\documentclass[10pt,epsf,preprint]{aastex}
\input psfig

\catcode`\@=11
\def\gsim{\ifmmode{\mathrel{\mathpalette\@versim>}}
    \else{$\mathrel{\mathpalette\@versim>}$}\fi}
\def\lsim{\ifmmode{\mathrel{\mathpalette\@versim<}}
    \else{$\mathrel{\mathpalette\@versim<}$}\fi}
\def\@versim#1#2{\lowerpaper2_9mar01.tex 2.9truept \vbox{\baselineskip 0pt \lineskip 
    0.5truept \ialign{$\m@th#1\hfil##\hfil$\crcr#2\crcr\sim\crcr}}}
\catcode`\@=12

\def\listitem{\par \hangindent=50pt\hangafter=1
     $\ $\hbox to 20pt{\hfil $\bullet$ \hfil}}

\def\puncspace{\ifmmode\,\else{\ifcat.\C{\if.\C\else\if,\C\else\if?\C\else%
\if:\C\else\if;\C\else\if-\C\else\if)\C\else\if/\C\else\if]\C\else\if'\C%
\else\space\fi\fi\fi\fi\fi\fi\fi\fi\fi\fi}%
\else\if\empty\C\else\if\space\C\else\space\fi\fi\fi}\fi}
\def\SP{\let\\=\empty\futurelet\C\puncspace}

\def\h-1{$h^{-1}$}

\def\void#1{{}}

\def\h1{$h^{-1}$}

\def\etal{et al.\SP}
\def\eg{e.g., \,}

\def\lsim{~\rlap{$<$}{\lower 1.0ex\hbox{$\sim$}}}
\def\gsim{~\rlap{$>$}{\lower 1.0ex\hbox{$\sim$}}}

\begin{document}

\title{Redshift-distance Survey of Early-type Galaxies: 
the ENEARc Cluster Sample  
\footnote{Based on observations at Complejo Astronomico El
Leoncito (CASLEO), operated under agreement between the Consejo
Nacional de Investigaciones Cient\'\i ficas de la Rep\'ublica
Argentina and the National Universities of La Plata, C\'ordoba and San
Juan; Cerro Tololo Interamerican Observatory (CTIO), operated by the
National Optical Astronomical Observatories, under AURA; European
Southern Observatory (ESO), partially under the ESO-ON agreement; Fred
Lawrence Whipple Observatory (FLWO); Observat\'orio do Pico dos Dias,
operated by the Laborat\'orio Nacional de Astrof\'\i sica (LNA); and
the MDM Observatory on Kitt Peak.} }

\author{M. Bernardi\altaffilmark{2},
M. V. Alonso\altaffilmark{3}, L. N. da Costa\altaffilmark{4,}\altaffilmark{5}, 
C. N. A. Willmer\altaffilmark{5,}\altaffilmark{6}}

\author{G. Wegner\altaffilmark{7},
P. S. Pellegrini\altaffilmark{5}, C. Rit\'e\altaffilmark{5},
M. A. G. Maia\altaffilmark{5}}

\affil{\altaffiltext{2}{The University of Chicago, 5640 South Ellis
Avenue, Chicago, IL 60637, USA}}

\affil{\altaffiltext{3}{Observatorio Astr\'onomico de
C\'ordoba,  Laprida  854, C\'ordoba, 5000, Argentina}}

\affil{\altaffiltext{4}{European Southern Observatory,
Karl-Schwarzschild Strasse 2, D-85748 Garching,
Germany}}

\affil{ \altaffiltext{5} {Departamento de Astronomia,
Observat\'orio Nacional, Rua General Jos\'e Cristino
77, Rio de Janeiro, R. J., 20921, Brazil}}

\affil{ \altaffiltext{6}{UCO/Lick Observatory, University of California,
1156 High Street, Santa Cruz,  CA 95064, USA}}

\affil{\altaffiltext{7}{Department of Physics \& Astronomy, Dartmouth
College, Hanover, NH  03755-3528, USA}}


\begin{abstract} 

This paper presents data on the ENEARc subsample of the larger ENEAR
survey of the nearby early-type galaxies. The ENEARc 
galaxies belong to clusters and were specifically 
chosen to be used for the construction of a $D_n-\sigma$ template.  
The ENEARc sample includes new measurements of spectroscopic and
photometric parameters 
(redshift, velocity dispersion, line index Mg$_2$, and 
the angular diameter $d_n$) 
as well as data from the literature.
New spectroscopic data are given for 229 
cluster early-type galaxies and new photometry is 
presented for 348 objects. Repeat and overlap
observations with external data sets are used to construct a final
merged catalog consisting of 640 early-type galaxies in 28
clusters. Objective criteria, based on catalogs of groups of galaxies
derived from complete redshift surveys of the nearby universe, are
used to assign galaxies to clusters. 
In a companion  paper these data are used 
to construct the template $D_n-\sigma$ distance
relation for early-type galaxies which has been used to
estimate galaxy distances and derive peculiar velocities for the ENEAR
all-sky sample.

\end {abstract}
\clearpage

\section{Introduction}

For many years cosmic flows have been a promising way to probe mass
density fluctuations on intermediate scales ($\lsim 100$\h1~Mpc) and
to obtain dynamical measures of the cosmological parameters. 
Unfortunately, compiling peculiar velocity data with the desired sky
coverage and full sampling is time-consuming, so most previous
cosmic flow studies have been based on sparse or
non-uniform catalogs, leading to conflicting interpretations.
Recently, new large redshift-distance surveys of
spiral (SFI; da Costa \etal 1996, Haynes \etal 1999a, 1999b) and early-type
galaxies (ENEAR; da Costa \etal 2000a, hereafter Paper~I; Wegner \etal
2002; Alonso \etal 2002) have been completed and some of the
unresolved issues are being re-addressed with 
these significantly larger and more uniform samples.

The ENEAR catalog of peculiar velocities 
described in Paper I is an all-sky catalog which includes a
magnitude-limited sample of early-type galaxies (ENEARm)
extracted from completed magnitude-limited redshift surveys, as well as a 
subsample of cluster galaxies which are fainter than the 
magnitude limit of the parent redshift surveys. 
The ENEAR catalog is a compilation of
observational parameters for $\sim 2000$ early-type galaxies. It
includes new photometric and spectroscopic data for $\sim 1500$
early-type galaxies in clusters and in the field as well as data
from previous work.
Our new photometric and spectroscopic measurements are 
presented in Alonso et al. (2002) and Wegner et al. (2002), respectively.  
Data from this catalog have 
already been used in some  previous analyses presented
in other papers of this series (Bernardi \etal 1998; 
Borgani \etal 2000; da Costa \etal
2000b; Nusser \etal 2001; Zaroubi \etal 2001). 

In conducting these analyses we have found it necessary operationally to
split the ENEAR catalog into two samples, ENEARm and ENEARc (see Paper~I).
The ENEARc is composed of galaxies from ENEARm that are in clusters,
but with the addition of
galaxies fainter than the original ENEARm magnitude limit 
using new observations
and data in the literature.

The purpose of the present paper is to
focus on the ENEARc cluster subsample; here we describe the assignment of
galaxies to clusters, the homogenization of our new data with those of 
previous work, and report the final combined photometric and spectroscopic
parameters of the cluster galaxies. 
The construction of the ENEARc catalog is needed in order to 
obtain an accurate template scaling relation 
(the $D_n-\sigma$ relation) which can be use  
to measure distances and peculiar velocities
for all the early-type galaxies in the ENEAR catalog. 
The $D_n-\sigma$ scaling relation, originally
introduced by Dressler \etal (1987), is essentially equivalent to the
more general Fundamental Plane (FP, Djorgovski \& Davis 1987) defined
by early-type galaxies.

The underlying assumption in estimating
distances is that these scaling relations are 
the same for field and cluster galaxies. 
For galaxies which are
at approximately the same redshift, so effects of evolution can be
neglected, the dependence of these scaling relations on cluster
properties has been investigated by several authors 
(\eg J{\o}rgensen, Franx, \& Kj{\ae}rgaard 1996; 
D'Onofrio \etal 1997; Pahre, Djorgovski, \& de Carvalho 1998; Scodeggio \etal
1998; Gibbons, Fruchter, \& Bothun 2001; 
Colless \etal 2001). (For how the FP depends
on cluster redshift see \eg J{\o}rgensen \etal 1999; Treu \etal 1999, 2001;
van Dokkum \etal 2001.)  Other issues that bear on the results are the
assignment of galaxies to clusters, the small number of observed
galaxies per cluster, which requires the use of all available cluster
data for the definition of a template distance relation, and the
possible biases of the parameters resulting from selection effects and
measurement errors. 

The data presented in this paper are used in Bernardi \etal (2002) to
derive the template-distance relation from which the galaxy distances and
the peculiar velocity field of the ENEAR sample were derived.  

The outline of this paper is as follows:
In Section~2 we describe the selection of clusters and
the criteria used for membership assignment.  In Section~3 we present
the new spectroscopic and photometric data and discuss the
corrections applied to our measurements and to those of
other authors which bring the available data into a common
system. Section~4, contains the final merged and
standardized ENEARc catalog of early-type cluster galaxies used to
construct the distance relation and peculiar velocity
analyses. A brief summary is given in Section~5.

\section {Cluster Sample}

\subsection{Selection}
\label{clusel}

A key requirement for constructing any  distance relation based on
cluster/group galaxies is to have a well defined procedure for identifying
bound systems and assigning  galaxies to them. With currently available 
optical, all-sky redshift
surveys (\eg Huchra \etal 1983, CfA1; Falco \etal 1999, CfA2; da Costa
\etal 1988, SSRS; da Costa \etal 1998, SSRS2; Santiago \etal 1995,
ORS) a significant improvement can be made 
on earlier work which justifies 
reexamining galaxy assignments to clusters. 

In
this paper we have adopted the following
procedure: Clusters within $\sim 10000$ kms$^{-1}$ were selected from
group catalogs derived by applying the objective group-finding
algorithm of Huchra \& Geller (1982) to all complete, magnitude-limited
redshift surveys currently available. For most of the sky ($\sim 6.5$
steradians) rich groups ($\gsim$ 15 members) were drawn from the CfA1
(Geller \& Huchra 1983) and SSRS (Maia, da Costa \& Latham 1989)
catalogs. At low galactic latitudes the catalogs were complemented by
groups identified in the Optical Redshift Survey (Santiago \etal
1995).  Over the fraction of the sky surveyed by the CfA2 and SSRS2
($\sim 4.2$ steradians, including more than a third of both galactic
caps), we used groups identified by Ramella, Pisani, 
\& Geller (1997) in
the CfA2 and Ramella \etal (2002) in the SSRS2, instead of those
identified in earlier shallower surveys. This
selection is not uniform over the sky because of the different
magnitude-limits and density contrast thresholds 
that had to be adopted in the group
identification. An attempt to minimize this effect and include
possible fainter members is discussed in Section~\ref{member}.

The final catalog contains
32 rich ($> 15$ members), 318 medium-size
(5-15 members) and 628 small ($< 5$ members) systems making a total of
978 groups. For
each group the combined catalog provides: the number of members;
center coordinates, heliocentric radial velocity, velocity dispersion,
and the physical size expressed by the
pair radius $R_p$ (Ramella, Geller, \& Huchra 1989) 
which is used below to
establish cluster membership. The physical parameters of the
groups are computed considering all morphological types.
Groups with at least 15 members were cross-identified with the Abell
and the ACO cluster catalogs (Abell 1958; Abell, Corwin \& Olowin
1989). All but three, consisting predominantly of spirals
were known clusters. Because of the
limiting magnitudes of the parent redshift surveys some faint
early-type galaxies, known to be in clusters, are missed by
this procedure. Fainter early-types with data in the literature were 
added to our compilation (See
Section~\ref{member}). 

To this list we added
five additional well-studied clusters: A539, AS639, and A3381
(J{\o}rgensen, Franx, \& Kj{\ae}rgaard 1995a, 1995b) and 
7S21 and A347 (Smith \etal 1997).  These 
clusters were excluded from our original list because
they are either located at low galactic latitudes, outside the regions
probed by the redshift surveys, or because the 
member galaxies are fainter than the limiting magnitude of these
redshift surveys.  A special procedure was also adopted to
handle the Centaurus cluster, which has two distinct components (Lucey
\& Carter 1988) but is identified as a single large system in the
group catalog. In this case, we assigned memberships based on the
observed redshift distribution along the line-of-sight.  The resulting
list of members for each system agrees well with Lucey \&
Carter (1988). The physical characteristics of these two groups,
hereafter Cen30 and Cen45, were computed after splitting the systems
(see Section~\ref{prop}).

\begin{figure}
\centering
\mbox{\psfig{figure=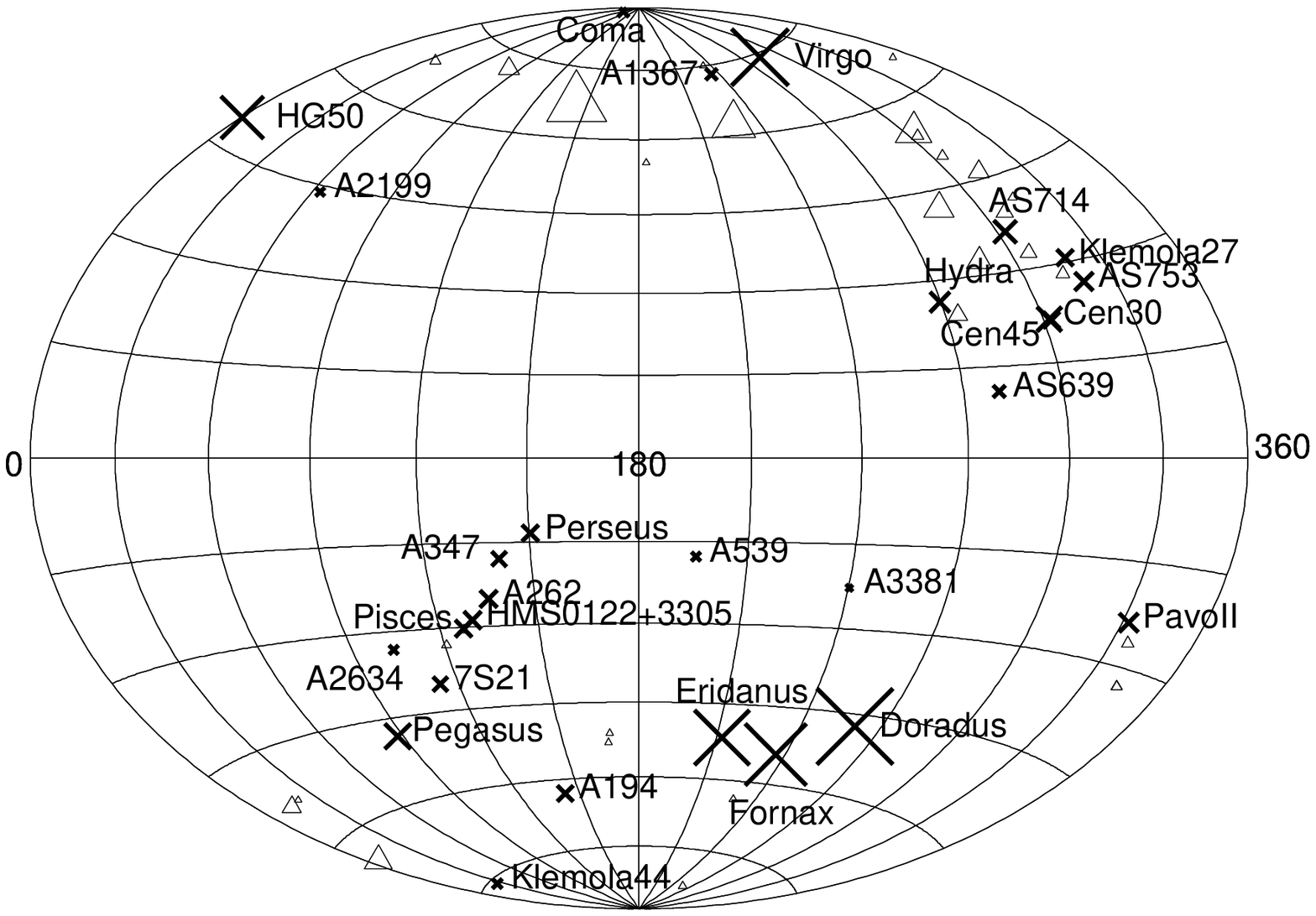,height=15truecm,bbllx=3truecm,bblly=5truecm,bburx=19truecm,bbury=23truecm }}
\caption{The spatial distribution in galactic coordinates of all the
58 cluster/groups with characteristics suitable for the determination
of the distance relation (see text). The 28 clusters used in the
present paper are shown as crosses. The size of the symbols is
inversely proportional to the cluster redshift.}
\label{fig:cludistr}
\end{figure}

From the cluster/group sample above we selected 58 groups containing  a
minimum of five early-type galaxies and at 
least 15 members of any morphological type, to provide a
reliable cluster velocity dispersion estimate. 
Figure~\ref{fig:cludistr} shows the projected
distribution in galactic coordinates of all the 58 selected clusters,
including the 28 clusters considered in the present paper. The apparent 
deficiency of clusters at low galactic latitudes may not be real. 
However the apparent underdensity of galaxies shown on the left side
of the plot at higher galactic latitudes can also be seen in
Figure 14 of Paper~I, which 
shows this for the ENEAR and SFI galaxies.  
We ultimately  focused on 
creating a large database of measurements in common with other 
authors, and to enlarge the sample of available galaxies in the 28 
previously studied clusters.

It is useful to refer to Figure~4 of Paper~I, which
compares the distribution of clusters on the sky with that of the
underlying galaxies.  Four clusters lie inside the
Perseus-Pisces (PP) region ($0^h <\alpha <4^h$ and $+20^o <\delta <+45^o$;
Smith et al. 1997).  The two dominant concentrations of galaxies, the
Great Attractor (GA) and Perseus-Pisces superclusters, are
indicated in that figure.  
Paper I also shows that in
our sample, the GA and PP superclusters produce a prominent
peak at $\sim 5000$ kms$^{-1}$ in the redshift distribution of
the clusters, indicating that 
our clusters probe the most prominent structures in the nearby 
universe.

\subsection {Properties of the Cluster Sample}
\label{prop}

The main physical characteristics of the 28 selected clusters are
given in Table \ref{tab:clusters}: the cluster name  
in column (1); columns (2) and (3) give the right ascension and
declination, as determined from the group finding algorithm or taken
from the references listed in column 9; column (4) is the heliocentric
radial velocity; column (5) the cluster velocity dispersion; 
column (6) the value of the radius $R_p$, whenever available (see
Section~\ref{member}); column (7) the number of early-type galaxies
with distances in this paper; column (8) indicates
clusters using only measurements obtained
by other authors; and column (9) references previous studies
of these clusters. Note that the cluster global parameters such as
redshift, velocity dispersion, and size were taken from the merged
group catalog described earlier. Comparing these redshifts
with those obtained using only the cluster early-types, we find
insignificant differences, the mean offset being $\sim
20$ kms$^{-1}$ and the scatter $\sim 110$ kms$^{-1}$. 
The cluster velocity dispersions listed in the table also agree well, 
with a mean difference of $\sim 40$ kms$^{-1}$ and a
scatter of $\sim 130$ kms$^{-1}$. In the latter case, we find that for
rich clusters, the values listed in the table are smaller with an 
offset of $\lsim 200$ kms$^{-1}$.

Poor sampling of the cluster region affects both the estimated
cluster velocity dispersion and the characteristic size
scale $R_p$. In some cases the coordinates of the 
groups/clusters had to be revised because closer inspection showed
that the group catalog was unable to separate neighboring clusters 
(\eg A2199/97, Cen30/45) or because groups based on shallow
surveys do not faithfully represent the center of the cluster after 
fainter members have been included (\eg Klemola 44).  
Groups found at the edge of a redshift survey tend to have 
their global parameters affected; in particular the size, as originally 
listed in the group catalog, can differ. For example, the group 
radius of Pavo~II depends on the density threshold.
This may also explain structure found in Doradus. These 
problems should be kept in mind when investigating correlations that 
depend on the global cluster parameters such as size and velocity 
dispersion.

\subsection {Membership Assignment}
\label{member}

As mentioned in Section \ref{clusel}, the group catalogs were derived
from redshift surveys with differing limiting magnitudes and selection
criteria. To make the distribution of our clusters/groups more uniform
across the sky we added other systems
in regions of the sky not covered by the redshift
surveys. Also, we have increased the number of cluster/group
members by including early-type galaxies observed by other
authors, below the magnitude limit of the redshift
surveys. We have also checked other authors' membership
assignments. A galaxy was assigned to a cluster if it
satisfied two conditions. The first condition is $d \le 1.5 R_p$,
where $d$ is the distance relative to the group center,
determined from the group finding algorithm, and $R_p$ is the group
pair radius (both are in $h^{-1}$ Mpc, where $h=H_0/100$ kms$^{-1}$
Mpc$^{-1}$). We fixed $R_p$ to 1 $h^{-1}$ Mpc for Cen45, not
identified by the finding algorithm, for Pavo~II at the edge
of the redshift survey, and for the five clusters added
from the literature. 

The second condition is $\mid cz -cz_{\rm cl}\mid
\le 1.5\sigma _{\rm cl}$ where $cz$ is the radial velocity of the
galaxy, $cz_{\rm cl}$ and $\sigma_{\rm cl}$ are the systemic velocity
and velocity dispersion of the group, respectively. This
second requirement does not precisely
represent the region within the caustic which marks the boundary of a
cluster in redshift space (\eg Kaiser 1987; Reg\H{o}s \& Geller 1989),
but the small number of galaxies per cluster and the 
caustic's dependence on 
$\Omega$ makes a more precise
determination untenable. 

To better approximate the real
assignment of galaxies to the cluster we also define a class of
``peripheral'' objects, the criteria
depending on the richness of the cluster.  For the richest clusters
(\eg Virgo and Coma), ``peripheral'' galaxies are those which satisfy
one of two conditions: $d \le R_p$ and $1.5 \sigma _{\rm cl} \le \mid
cz- cz_{\rm cl}\mid \le 3\sigma _{\rm cl}$, or $1.5 R_p\le d \le 3
R_p$ and $\mid cz- cz_{\rm cl}\mid \le \sigma_{\rm cl}$, while for
clusters with fewer members the second condition becomes $1.5 R_p \le
d \le 2 R_p$.  These conditions are intended to represent the region
in redshift-space occupied by cluster members. Applying these criteria,
increases the sample of the early-type galaxies 
by about 20\%. When two clusters are close to each other as
in the cases of A2199/97 and Cen30/45 ambiguous cases remain. The
impact of these objects on the distance relation is
discussed in more detail in Bernardi \etal (2002).

\begin{figure}
\centering
\mbox{\psfig{figure=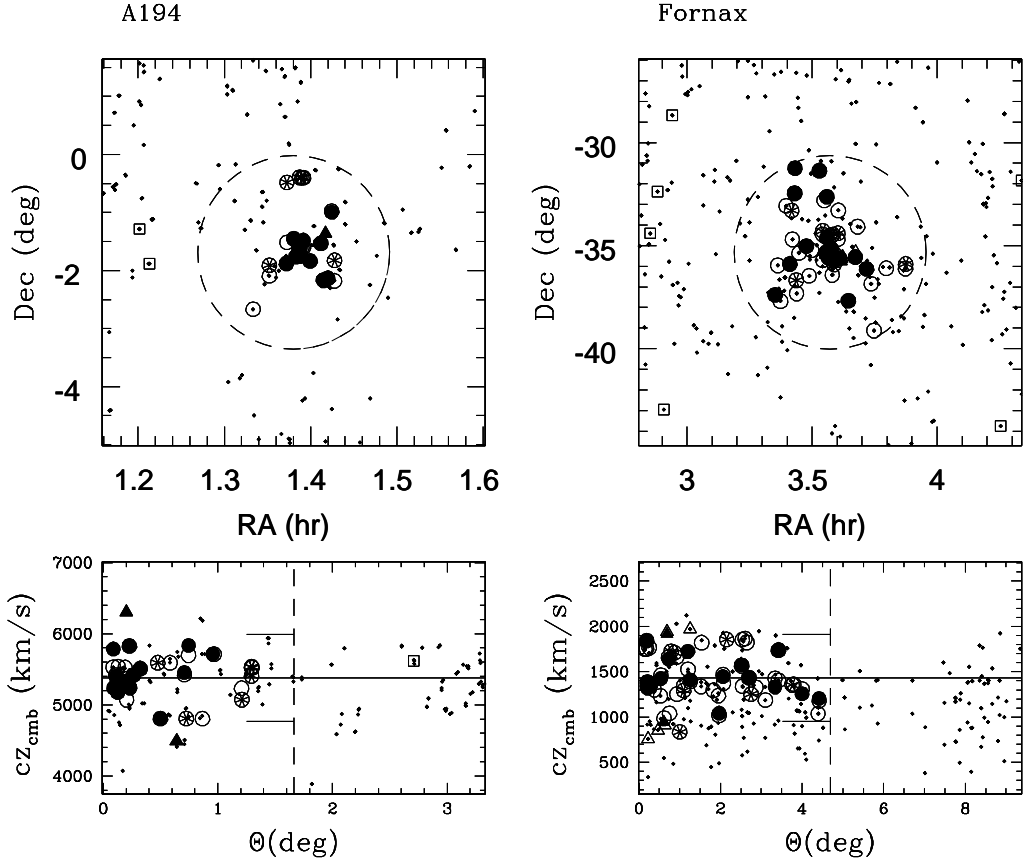,height=15truecm,bbllx=3truecm,bblly=6truecm,bburx=19truecm,bbury=24truecm}}
\caption{Examples of
the distribution of galaxies, in the equatorial coordinate
system (upper panel), and the radial velocity versus angular distance
from the cluster center (lower panel) for two clusters. 
Shown are:  late-type
field galaxies taken from the available redshift surveys (small dots)
and early-type field galaxies in the ENEAR catalog (open squares);
cluster early-type galaxies with (filled and skeletal circles) or
without measured distances (open circles).  The skeletal symbol
represents ``discarded'' galaxies (see the text and Table~2 for
details); and ``peripheral'' galaxies
with (filled triangles) or without measured distances (open triangles).}
\label{fig:clu}
\end{figure}

Figure~\ref{fig:clu} shows examples of the resulting projected 
distribution of objects in and around two of the clusters 
(similar plots for the other clusters are available;
interested readers should contact the first author of this paper
directly).
The dashed circle 
corresponds to the angular size of $1.5 R_p$ at the cluster redshift.
Some systems are well sampled by the
available redshift surveys (\eg Coma, A1367, Virgo, Fornax, Eridanus,
Doradus), while at low galactic latitude clusters/groups are only covered by
the shallow ORS. The density of
galaxies abruptly changes south of $-40^\circ$, the southern
limit of the SSRS2. From these projected 
distributions, we also found that some 
group centers, in the group catalogs, are
not good estimates. This is the case, \eg for Pisces,
HMS0122+3305, A2199, and the nearby groups such as Eridanus and
Doradus, but correcting the central positions leaves
the galaxy assignments unaltered. We also also find that Klemola~44 may be a
superposition of two systems. 

Overall $\lsim 90\%$
of the early-type galaxies assigned to these groups have distances.
There are 733 early-type galaxies assigned to the 28 selected
clusters/groups (hereafter the ENEARc catalog), of which 640 galaxies have
velocity dispersions and $d_n$ measurements, where $d_n$ is the
apparent diameter at a given isophotal level, as required to build the
$D_n-\sigma$ relation.  Of these, 495 satisfy the most stringent
membership criteria, whereas 145 are ``peripheral'' cluster objects.
According to our adopted criteria, 15 galaxies assigned
to the selected cluster sample by previous authors are non-members.  

These non-member galaxies are listed in Table~\ref{tab:litwrong},
where: column (1) is the name of the galaxy; columns (2)
and (3) are the equatorial coordinates; column (4) is the heliocentric
redshift; column (5) is the total magnitude $m_{\rm B}$; column (6) is
the name of the cluster to which the galaxy is assigned; column (7) is
the pair radius $R_p$ of the cluster; column (8) is the cluster
velocity dispersion; column (9) is the projected distance of the
galaxy from the cluster center, computed using the angular separation
and the cluster redshift; column (10) is the difference between
the galaxy and cluster redshifts; and column (11) references 
previous work. The information on the individual galaxies and
groups comes from the literature and the original group catalogs,
respectively. Note that Willmer \etal (1991) had already pointed out
that [WFC91]~056 and ESO~384G037 are interlopers in AS753, while
J{\o}rgensen (1997) noted that D75 and [WFC91]~039 were not members of
A194 and AS753, respectively. The galaxies in Table~\ref{tab:litwrong}
are, therefore, excluded from the ENEARc sample. 

\begin{figure}
\centering
\mbox{\psfig{figure=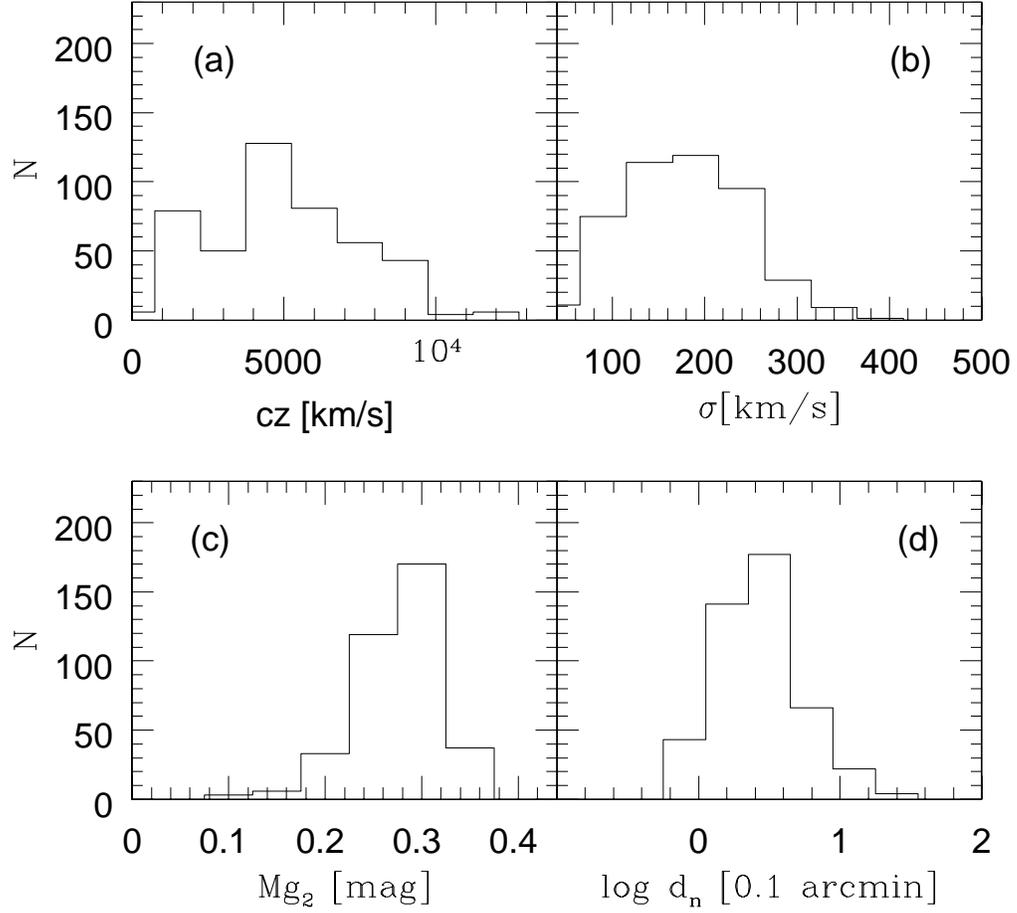,height=12truecm,bbllx=0truecm,bblly=10truecm,bburx=19truecm,bbury=22truecm }}
\caption{The distribution of spectroscopic and photometric parameters
for the ENEARc galaxies: upper panels (a) redshift and (b) velocity
dispersion; lower panels (c) Mg$_2$ line index and (d) the photometric
parameter log~$d_n$.}
\label{fig:histgalpar}
\end{figure}

Figure~\ref{fig:histgalpar} summarizes the measured parameters for
early-type cluster galaxies. This figure shows the distribution of:
redshift (panel a); $\sigma$ (panel b); Mg$_2$ index (panel c); and
the photometric parameter $d_n$ (panel d).  The data  
include both our  measurements and those of other
authors calibrated onto a common system as described
in the next section.

\section{Homogenization of the Measurements}

In addition to our all-sky ENEAR redshift-distance survey, over the years we
have gathered a large number of spectroscopic and photometric
observations of early-type galaxies in clusters with the following
three goals in mind: 1) increase the number of galaxies in clusters
to improve the statistical accuracy of
distance relations; 2) measure many galaxies
observed by other authors, in order to scale the available data into a
common system and estimate our external errors; 3) obtain 
repeat observations to estimate our internal errors on a
run-by-run basis. Repeat measurements also
provide a way to eliminate disparate measurements and improve the
statistical error per galaxy.
Since a non-negligible number of galaxies, mostly in
clusters, had modern high-quality measurements available in the
literature, we concentrated our efforts in clusters already studied by
other authors.  

In the following sections we describe the procedures adopted for
homogenizing and combining the spectroscopic and photometric
parameters which come from various sources.

\subsection {Spectroscopic Data}

Details of the observations, data reductions, parameter and error estimates
obtained by the ENEAR team are presented in Wegner \etal (2002). 
The velocity dispersion was measured in the interval $4770-5770$~\AA.  
Therefore, our estimates do not include the effects of the NaD line which, 
because of interstellar absorption, could affect the velocity dispersion 
measurement (\eg Dressler 1984).
It is known that for any given $S/N$ and instrumental resolution, there is 
a lower limit on
the velocity dispersion measurable without introducing significant bias (\eg
Bender 1990).  The instrumental resolution of the setups used by the ENEAR team
range from 70~kms$^{-1}$ to 100~kms$^{-1}$. We tested the reliability of the
low velocity dispersion measurements by using: a) internal repeated
observations, and b) simulations of synthetic noisy ``galaxy'' spectra (Wegner
\etal 2002).  The scatter (~15~\%) given by internal comparisons at low
velocity dispersions ($< 70$~kms$^{-1}$) is comparable with the errors
(10~\%-15~\%) associated with those measurements.  Simulated ``galaxy''
spectra, obtained by broadening the spectrum of a template star by Gaussians of
different velocity dispersion and by adding different noise, show that for
spectra with $S/N > 20$ per pixel the velocity dispersion can be measured
reliably down to a value of $\sim 65$~\% of the instrumental resolution.  These
results show that our random errors are reliable and that our velocity
dispersion measurements are not biased---even for values which are smaller than
the instrumental resolution.  However, the above tests do not take into account
about possible systematic effects due to \eg template and galaxy mismatches or
continuum subtraction problems. 

In this section we describe the homogenization procedure for the ENEARc data
which include our new measurements as well as data from the literature.  
The first step in homogenizing the  
velocity dispersions and line indices is to define a
``fiducial'' system to which all other measurements can be compared
and converted. To achieve this we first brought our measurements onto
the same internal system. The ENEAR observations were chosen as this
standard system because they are a large ($\sim 2000$) homogeneous set of
measurements, which was designed to extensively overlap with other
samples and our own individual observing runs. 
The ENEAR observations were brought into a consistent system as  
described by Wegner \etal (2002).
Briefly, the following procedure was used: 
we determined a mean offset, $\Delta$, between each galaxy in
a given run and all the other runs. We then averaged over all galaxies in a 
run, weighting by the number of available pairs which yields $\Delta$
for that run. We then subtracted $\Delta$ from data in each run and repeat 
the process until no run has an offset above a predefined tolerance. 
We found no evidence for a more complicated correction for the data.
The internal comparisons of the final redshift,
velocity dispersion, and the Mg$_2$ index were found to be in good
agreement, showing no systematic trends, and only small zero-point
offsets.

The mean redshift offset is $\lsim 25$~kms$^{-1}$ with a scatter of
$\lsim 40$~kms$^{-1}$; in the case of the velocity dispersion, $\Delta
\log{\sigma}$ is always less than $0.025$ and has a
scatter of $\lsim 0.060$~dex. The mean $\Delta \log{\sigma}$ offset is
$0.012\pm 0.008$~dex.  Finally, the line index Mg$_2$ has a mean
zero-point offset of $\lsim 0.015\pm 0.006$~dex with a scatter of
$\sim 0.020$~dex. The velocity dispersions and the values of the
Mg$_2$ index were aperture-corrected to $2r_{norm} = 1.19 h^{-1}$~kpc
(Wegner \etal 2002).

The second step,
after calibrating our measurements to a consistent system, was to
convert published data to
our reference system. An aperture correction
consistent with our raw data was made before calibrating them
to our ``fiducial'' system by a zero-point shift, 
using a procedure similar to the one
above. Zero-point corrections were only applied for those data sets for
which the offset was larger than its error. Table~\ref{tab:corrsp}
summarizes the results of this comparison of our data with those of
other authors: in column (1) the source of published data; in column
(2) the number of velocity dispersion measurements involved in the
comparison; in column (3) the offset applied to each set of velocity
dispersion measurements; in column (4) the rms value of the
differences in the comparison; columns (5)-(7) show the same for the
Mg$_2$ line index. All differences are computed as
``our-minus-literature''. The comparisons listed in the table include
only cluster galaxies except for the 7S sample; for that sample
we also compared the measurements of field galaxies. This was done
to increase the accuracy of the comparison.

\begin{figure}
\centering
\mbox{\psfig{figure=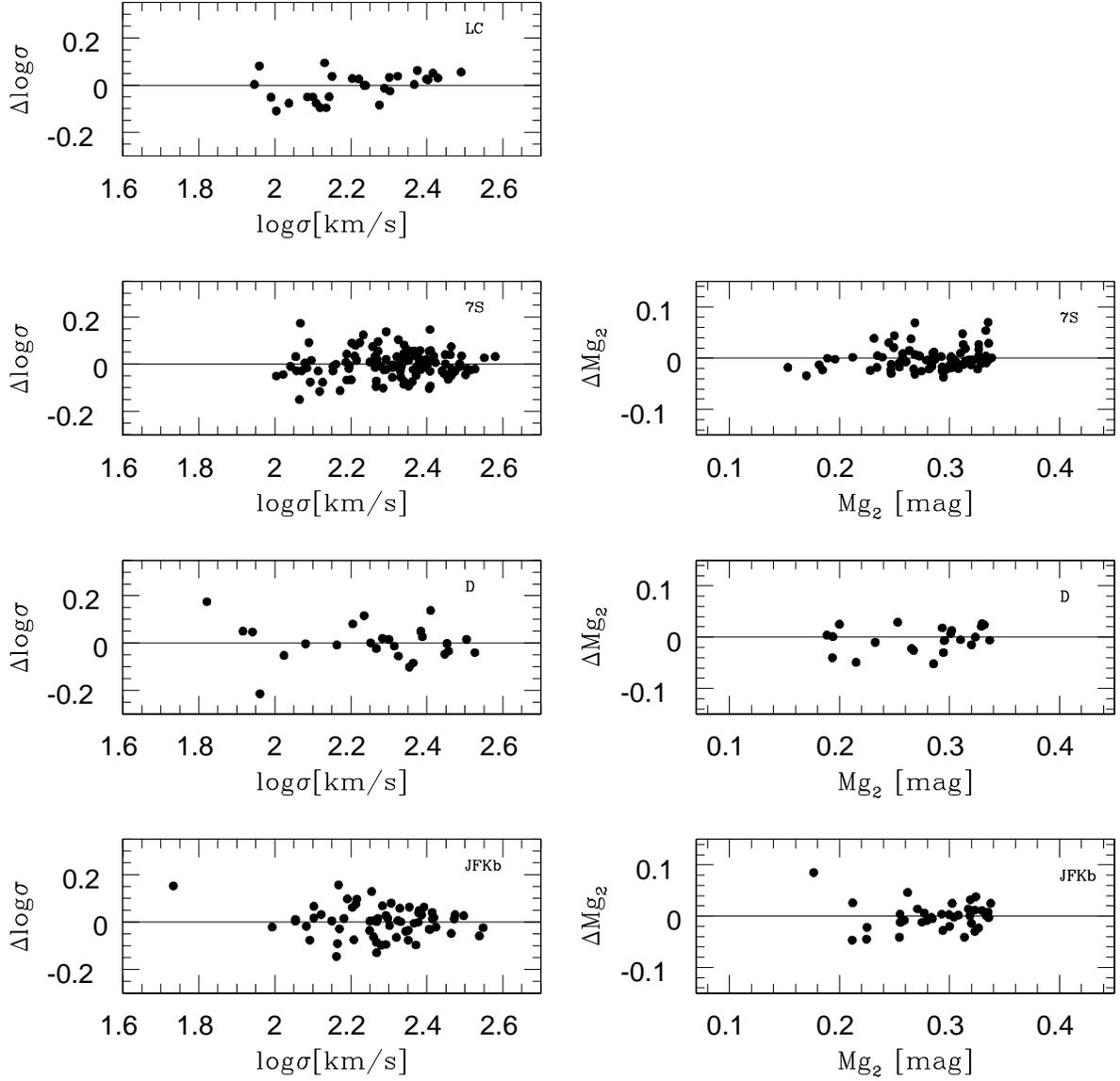,height=14truecm,bbllx=2truecm,bblly=5truecm,bburx=19truecm,bbury=22truecm }}
\caption{Comparison of our measurements with those of other authors.
Different panels show the comparison for each individual source: Lucey
\& Carter (1988) (LC); Faber et al. (1989) (7S);  Dressler (1987) and
Dressler, Faber, \& Burstein (1991) (D); 
and J{\o}rgensen et al. (1995b) (JFKb).}
\label{fig:compsp_lit1}
\end{figure}

\begin{figure}
\centering
\mbox{\psfig{figure=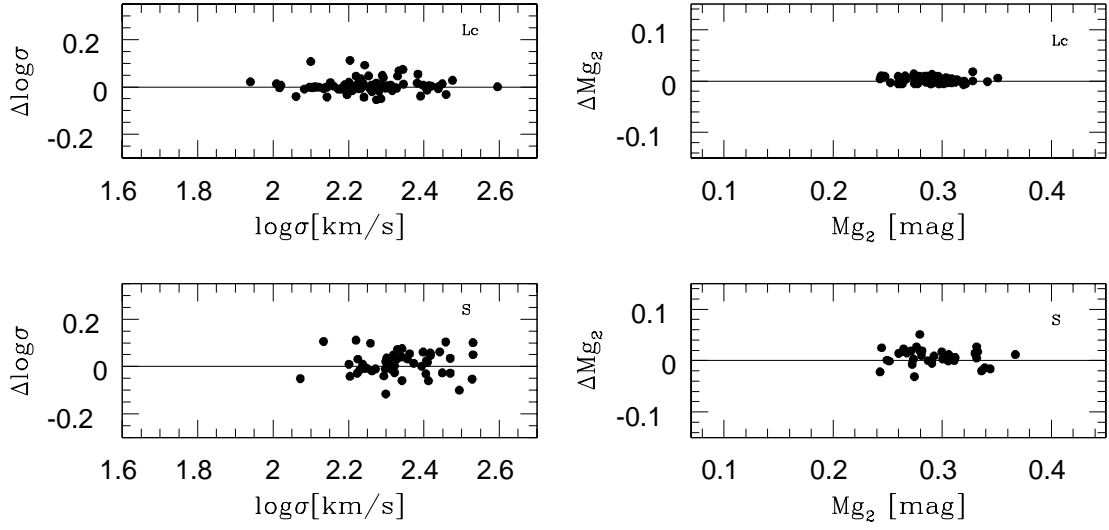,height=14truecm,bbllx=3truecm,bblly=6truecm,bburx=19truecm,bbury=24truecm }}
\caption{Comparison between measurements from different authors for
which no direct calibration was possible (Lucey et al. 1997 (Lc) and
Smith et al. 1997 (S)) after bringing all measurements into 
a common system.  }
\label{fig:compsp_lit2}
\end{figure}

In total we have 248 measurements of the 
velocity dispersion and 156 of the Mg$_2$ index in common with the
literature. The external comparisons reported in the table show that
the external offsets are, in general, comparable to the mean offset
found from our internal comparisons. The measurements obtained by
other authors were converted to our ``fiducial'' system by adding the
offsets listed in Table~\ref{tab:corrsp}. In order to evaluate the
final merged and standardized catalog of early-type galaxies a
comparison has been made between different (calibrated) measurements
of the same galaxy. The results are shown in
Figure~\ref{fig:compsp_lit1} for those datasets for which direct
comparisons are available. The data sets of Lucey \etal (1997) and
Smith \etal (1997) have very few galaxies in common with our
sample. Therefore, for these two data sets we had to rely on an
indirect calibration. This was carried out by using the overlap
between the data of these two papers with those of the literature 
converted directly into our system. Figure~\ref{fig:compsp_lit2}
shows the result of this indirect calibration.

\begin{figure}
\centering
\mbox{\psfig{figure=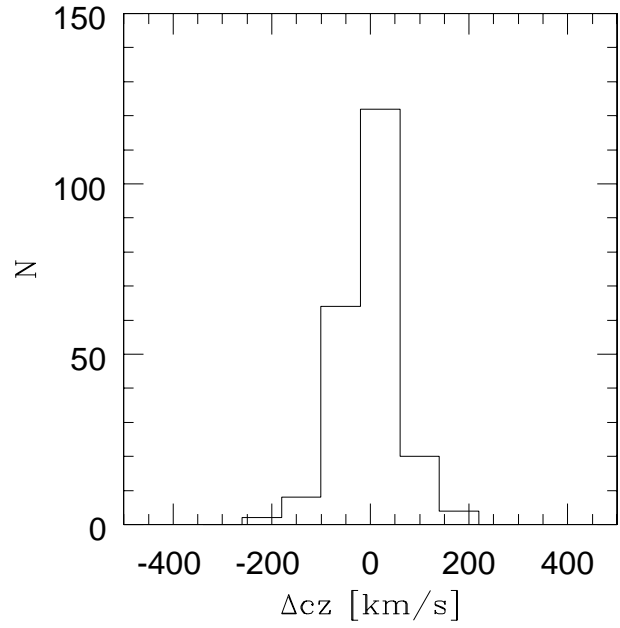,height=11truecm,bbllx=-2.5truecm,bblly=10truecm,bburx=19truecm,bbury=22truecm}}
\caption{The distribution of the redshift difference between our
measurements and those from the literature.}
\label{fig:histczcl}
\end{figure}

Figure~\ref{fig:histczcl} shows a histogram of the difference between
our redshifts and the literature.
The mean and rms scatter of the distribution are $1\pm 8$
kms$^{-1}$ and $69$ kms$^{-1}$, respectively. Since the typical error
in the radial velocity is of the order of $\sim 50$~kms$^{-1}$ we
conclude that no correction is required for the redshift
measurements.

The spectroscopic data in the ENEARc catalog are summarized
in Table~\ref{tab:speobcl} which gives: in column (1) the source of
measurements; in column (2) the number of new measurements obtained by
our team; in column (3) the
number of galaxies for which spectroscopic parameters were measured;
in column (4) the number of galaxies for which there is only one
source; in columns (5)--(7) the same information for the Mg$_2$ line
index. In estimating the contribution of each
author we tried to consider only independent
measurements, although this is not always possible because some
authors combined their new data with
previous work. The number of measurements in the table 
reported for the authors refers only to galaxies in the
ENEAR database (see Paper~I).

Table~\ref{tab:speobcl} shows that we obtained 338 measurements of
the velocity dispersion for 229 galaxies and similar numbers for the
Mg$_2$ index; about one-third of our observations have
repeats. Using these data, 
we obtained new distances and Mg$_2$ indices for
$\sim 90$ galaxies, thereby enlarging the sample of early-type
galaxies that can be used in the construction of the $D_n-\sigma$
relation. About 55\% of the cluster sample still relies on 
single velocity dispersion measurements but most of these
are now based on high-quality modern observations.

\subsection{Photometric Data}
\label{photometry}

Details of the photometric observations are given in Alonso \etal (2002).
The same homogenization
procedure described above was adopted for the photometric
parameter $d_n$. Following the 7S, $d_n$ is the angular diameter of a
circular aperture within which the average integrated surface
brightness of a galaxy is equal to a specified value. In our case the
angular diameter $d_n$ is measured at the $R$-band isophotal level of
$\mu_ R=19.25$ mag arcsec$^{-1}$.  This value roughly corresponds to
the value adopted by the 7S in the $B$-band, assuming a mean color of
$(B-R)$=1.5~mag. The
light profiles were corrected for the effects of seeing (\eg Saglia
\etal 1997), flux calibrated and corrected for Galactic extinction
(Burstein \& Heiles 1984). Alonso \etal 2002 show that, for the 
galaxies in our sample, the Burstein \& Heiles extinction corrections 
agree well with the values derived 
using the extinction maps of Schlegel, Finkbeiner \& Davis (1998).
The K-correction and cosmological surface brightness
dimming correction were also applied. 

As in the case of the spectroscopic parameters,
a homogeneous dataset is required. This is particularly important for
combining photometric parameters since these come from a variety of sources,
many using different filters.
To calibrate the photometric parameter $d_n$ to the ``fiducial''
system, we first checked the internal consistency of our data by
comparing the surface brightness profiles of galaxies for which we
have more than one observation.  For these objects, the dispersion
among different measurements was found to be small ($\sim 0.05$
mag/arcsec$^{2}$ over an interval of typically 3 magnitudes), showing
that our reduction and calibration procedures lead to uniform
results (see Figure~3.6, 3.7, and 3.8 in Bernardi~1999). We also 
compared our measures of the surface brightness profiles
with those in previous works. This was done to estimate if
differences in the zero-point of the photometric calibration, or 
variations in the filters/colors used by different sources could
contribute to the differences observed in the $d_n$ parameter. 
We found that the comparisons of our surface brightness profiles
with those of other sources do not show any systematic gradient
or statistically significant offset (see Figure~3.4 in Bernardi~1999). 
These results justify our choice of applying simple offsets in homogenizing
the $d_n$ parameter. 
As before, our observations were used to define the reference
system; we brought the measurements to a common system by
minimizing the 
mean differences in the $d_n$ derived from galaxies observed
in more than one run.  The required corrections to bring the
measurements into a common system were relatively small: $\Delta
\log{d_n} \lsim 0.010\pm~0.004$~dex with a rms scatter of
$\lsim~0.022$~dex (Alonso \etal 2002).

\begin{figure}
\centering
\mbox{\psfig{figure=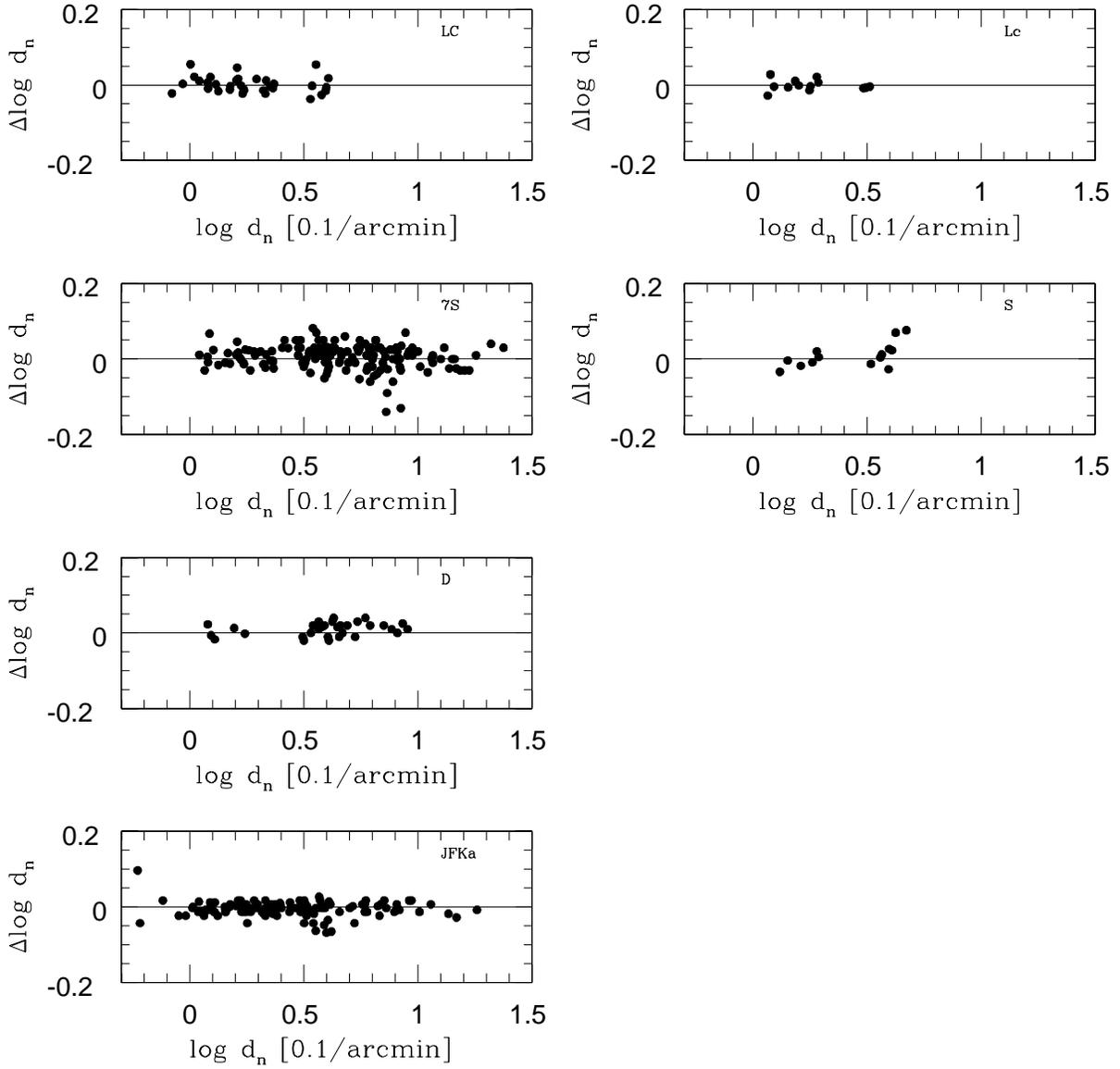,height=14truecm,bbllx=2truecm,bblly=5truecm,bburx=19truecm,bbury=22truecm }}
\caption{Comparison of our measurements of $\log{d_n}$ with those 
of other authors.  Different panels show this comparison for 
individual sources: Lucey \& Carter (1988) (LC);
Faber et al. (1989) (7S);
Dressler (1987) and Dressler et al. (1991) (D); 
J{\o}rgensen et al. (1995a) (JFKa);
Lucey et al. (1997) (Lc); and Smith et al. (1997) (S).} 
\label{fig:compdn_lit}
\end{figure}

The measurements available in the literature were calibrated to our
internal reference system by applying the offsets listed in
Table~\ref{tab:corrph}, if the offset is larger than its error. This
table summarizes the results of the comparison between our data and
those of other authors listing: in column (1) the source of published
data; in column (2) the number of measurements involved in the
comparison; in column (3) the difference between our measurements and
those in the literature; and in column (4) the rms value of these
differences.  There are 379 $d_n$ measurements in common with
the literature. The required corrections are small and the
scatter compares with that obtained from our internal comparisons,
suggesting that one can safely combine the measurements from different
data sets. These comparisons are shown in Figure~\ref{fig:compdn_lit}
after each individual data set was calibrated to the fiducial
system. The agreement is good, with the
dispersion being generally smaller than 0.02~dex. Furthermore, there
is no evidence of systematic trends, justifying our use of single
offsets to bring all of the published data into a common system.
As for the spectroscopic parameters, the comparison with the 7S sample 
also includes field galaxies. 

The photometric data assembled for the ENEARc sample are summarized in
Table~\ref{tab:phoobcl} where we list: in column (1) the source of
measurements; in column (2) the passband in which the data were
measured; in column (3) the number of new measurements including
repeat observations; in column (4) the number of galaxies with
measured photometric parameters; and in column (5) the number of
galaxies for which there is only one source.
We have measured 508 $d_n$ for a total of 348 ENEARc galaxies, of
which 117 had no previous measurements. 
About $48$\% of the cluster sample still relies on single
measurements, making the cross-comparison between different sources
extremely important. 

\section {The ENEARc Catalog}
\label{data}

As a final result, 
we have assembled a sample of 640 early-type galaxies in 28 clusters
with redshift, velocity dispersion, $d_n$ measurements and, whenever
possible, the Mg$_2$ line index using the membership criteria described
above. Of these galaxies, 495 are considered
cluster members and 145 ``peripheral'' objects.  All 640 objects were
individually inspected and 188~galaxies were removed due to 
peculiarities and/or problems on their images and/or 
their spectra (\eg residual
contamination from nearby galaxies or stars; presence of spiral arms
or bar; dust lane; high $D/B$; emission lines; low $S/N$) 
that could affect the measurement of their photometric and
spectroscopic parameters.  

Table~7  tabulates the main
measured parameters for the 452 galaxies suitable for constructing the
$D_n-\sigma$ relation and  Table~8 contains the 
188 cluster galaxies eliminated from
further consideration, respectively.  For each cluster these tables
give : in column (1) the name of the galaxy; in columns (2) and (3)
the 1950.0 equatorial coordinates; in column (4) the morphological $T$
type following Lauberts \& Valentijn (1989); in column
(5) the total $B$-band magnitude $m_{\rm B}$ taken from the
literature; in column (6) the number of redshift and velocity
dispersion measurements obtained from our new data; in column (7) the
number of redshift and velocity dispersion measurements available in
the literature; in columns (8) and (9) the heliocentric redshift and
its error; in columns (10) and (11) the logarithm of the velocity
dispersion measurement and its error; in columns (12) and (13) the
number of Mg$_2$ line index measurements available from our new data
and from the literature; and in columns (14) and (15) the value and
error of the Mg$_2$ line index. Similarly, columns (16)-(19) give the
same information for the photometric parameter $\log d_n$ ($d_n$ in
0.1~arcmin). In addition, Table~8 gives (column~20) a
reference to the Notes to the table indicate
the reason(s) for removing the galaxy from the ENEARc
sample.  The parameters listed in these tables are
combined values obtained from the error-weighted mean of the individual
measurements using a $3\sigma$-clipping.

\section {Summary}

In this paper we present spectroscopic (redshift, velocity dispersion,
and Mg$_2$ index) and photometric ($d_n$) data for 640 galaxies in 28
clusters comprising our ENEARc catalog.  
The assignment of galaxies to groups and clusters was based
on a compilation of objectively identified groups derived from
complete redshift surveys of the nearby universe. Roughly
2\% of the galaxies were previously assigned to clusters erroneously.
The data presented here are a combination of 338 new spectroscopic
measurements of 229 galaxies, and 508 new $R$-band images of 348
galaxies, in addition to those taken from the literature.  The
large number of galaxies in common with other authors, 
permits all data to be 
calibrated into one reference system. Bernardi
\etal (2002) use 452 galaxies of ENEARc to determine the $D_n-\sigma$ relation
used for ENEAR.

In a forthcoming paper we intend to also measure the FP parameters for
the ENEARc galaxies in order to build a template relation for nearby
clusters which will serve as a reference for similar studies at high
redshift. It is important to point out that the present sample can be
significantly expanded by using currently available wide-field imagers
and multi-object spectrographs such as 6dF
(http://www.aao.gov.au/ukst/6df.html) and it may be well worth the
effort.

\acknowledgments{The authors would like to thank the referee
for all the helpful comments and all of those who have
contributed directly or indirectly to this long-term project.  Our
special thanks to Ot\'avio Chaves for his many contributions over the
years. We would also like to thank D. Burstein. M. Davis, A. Milone,
M. Ramella, R. Saglia, and B. Santiago for useful discussions and
input.  MB thanks the Sternwarte M\"unchen, the Technische
Universit\"at M\"unchen, ESO Studentship program, and MPA Garching for
their financial support during different phases of this research.  MVA
thanks CNPq for different fellowships at the beginning of the project
and the CfA and ESO's visitor programs for support of visits.  MVA is
partially supported by CONICET and SecyT. LNdC would like to extend
his special thanks to David W. Latham who played a pivotal role at the
early stages of this project. GW is grateful to the Alexander von
Humboldt-Stiftung for making possible a year's stay at the
Ruhr-Universit\"at in Bochum, and to ESO for support for visits to
Garching.  Financial support for this work has been given through
FAPERJ (CNAW, MAGM, PSSP), CNPq grants 201036/90.8, 301364/86-9
(CNAW), 301366/86-1 (MAGM); NSF AST 9529098 and 0071198 (CNAW); ESO Visitor grant
(CNAW). PSP and MAGM thank CLAF for financial support and CNPq
fellowships. Most of the observations carried out at ESO's 1.52m
telescope at La Silla were conducted under the auspices of the
bi-lateral time-sharing agreement between ESO and MCT/Observat\'orio
Nacional. We are grateful to the anonymous referee whose detailed
comments greatly improved this paper.}

{}

\clearpage


\clearpage


\normalsize
\begin{table}
\begin{center}
\caption{The cluster sample}
\vspace*{10truemm}
\scriptsize
\begin{tabular}{lcrccccccr}
 \tableline  \tableline
\label{tab:clusters}
Name &$\alpha$ & $\delta$\ \ \ \ \ & $cz_{\rm hel}$&
$\sigma_{\rm cl}$ & R$_p$ & N$_{\rm gal}$ & & Reference \\
 & (1950) & (1950)\ \ & kms$^{-1}$ & kms$^{-1}$ & $h^{-1}$ Mpc & & & \\
(1) & (2) & (3)\ \ \ \ & (4) & (5) & (6) & (7) & (8) & (9)\\
 \tableline
\\[-2mm]
{\small \bf Northern hemisphere} \\
\\[-2mm]
  7S21           & 00:18:36 & 22:00:29 & \ 5840& 285 & -     & \  7 &*&S, G \\ 
  Pisces         & 01:06:21 & 32:16:19 & \ 5011& 358 &  0.837&   24 & &F, S\\
  HMS0122+3305   & 01:20:33 & 33:22:22 & \ 4884& 540 &  1.065&   11 & &S \\
  A262           & 01:50:00 & 35:56:19 & \ 4982& 450 &  1.130& \  9 &&F, S, W\\
  A347           & 02:21:36 & 41:25:00 & \ 5519& 768 & -     & \  8 &*&S, G \\
  Perseus        & 03:15:18 & 41:19:58 & \ 4967& 550 &  1.110&   30 & &D, F, S\\
  A539           & 05:13:55 & 06:24:00 & \ 8646& 701 &  -    &   19 &*&JFK\\
  A1367          & 11:41:41 & 20:15:25 & \ 6486& 760 &  1.043& \  8 & &F, HG\\
  Virgo          & 12:28:02 & 12:22:55 & \ 1101& 685 &  0.846&   81 & &F, HG \\
  Coma           & 12:56:01 & 28:07:18 & \ 7010& 895 &  1.925&   97 &&D, Da, F, JFK, Lc, S, W \\
  HG50           & 15:02:35 & 02:07:39 & \ 1707& 249 &  0.451& \  8 & &F, HG\\
  A2199          & 16:25:48 & 39:37:48 & \ 9075& 731 &  1.466&   20 &*&F, Lc, S, W\\
  Pegasus        & 23:17:57 & 08:04:21 & \ 3567& 405 &  0.444& \  5 & &F\\
  A2634          & 23:35:45 & 26:38:39 & \ 9318& 814 &  0.917&   14 &*&F, Lc, S, W\\
\\[-2mm]
{\small \bf Southern hemisphere} \\
\\[-2mm]
  A194           & 01:24:37 &-01:41:10 & \ 5380& 408 &  1.042&   21 & &F, LC, JFKa, JFKb \\
  Fornax         & 03:34:19 &-35:19:27 & \ 1433& 320 &  0.781&   25 & &D, F, HG, MDL\\
  Eridanus       & 03:35:25 &-20:38:24 & \ 1618& 236 &  0.682&   22 & &HG, Wa\\ 
  Doradus        & 03:59:05 &-51:16:59 & \ 1114& 245 &  1.300&   10 & &HG, JFK, MDL \\
  A3381          & 06:08:06 &-33:34:59 &  11381& 372 &  -    &   13 &*&JFK\\
  Hydra          & 10:34:19 &-27:26:53 & \ 3720& 555 &  0.867&   50 & &F, JFK, LC \\
  AS639          & 10:38:24 &-45:55:59 & \ 6246& 456 &  -    &   10 &*&JFK\\
  Cen45          & 12:46:48 &-40:56:19 & \ 4650& 350 &  -    &   10 &*&F, HG, LC\\
  Cen30          & 12:44:12 &-40:44:19 & \ 3030& 658 &  1.385&   36 & &Db, F, HG, LC\\
  AS714          & 12:48:17 &-26:17:27 & \ 3258& 215 &  0.450& \  8 & &A\\
  Klemola27      & 13:46:11 &-30:22:03 & \ 4611& 382 &  0.670&   18 & &Db, JFK, Wb\\
  AS753          & 13:58:31 &-33:40:08 & \ 4167& 401 &  0.979&   27 & &Db, JFK, Wb\\
  PavoII          & 18:43:49 &-63:28:48 & \ 4286& 284 &  -   & \ 17 & &F, HG, LC\\
  Klemola44       & 23:44:52 &-28:22:17 & \ 8457& 375 &  1.034&   32 & &F, JFK, LC\\
 \tableline
\end{tabular}
\end{center}
\footnotesize
Notes. --- Asterisks in column (8) denote clusters that were not observed by us; 
data for these clusters come entirely from other authors.\\ 
References. ---
A: Abell et al. (1989);
D: Dressler et al. (1987);
Da: Dressler (1987);
Db: Dressler et al. (1991);
F: Faber et al. (1989);
G: Gibbons \etal (2000);
HG: groups form Huchra \& Geller. (1982);
JFK: J{\o}rgensen et al. (1996);
LC: Lucey \& Carter (1988); 
Lc: Lucey \etal (1997);
MDL: groups from Maia et al. (1989);
S: Smith et al. (1997);
W: Wegner et al. (1999);
Wa: Willmer et al. (1989); and
Wb: Willmer et al. (1991).
\end{table}

\clearpage

\normalsize
\begin{table}
\begin{center}
\caption{Galaxies excluded by our cluster membership assignment}
\vspace*{10truemm}
\scriptsize
\begin{tabular}{lcrcclccccl}
 \tableline  \tableline
\label{tab:litwrong}
Galaxy & $\alpha$ & $\delta$\ \ \ \ \ & $cz_{\rm hel}$ & $m_{\rm B}$ &
Cluster & R$_p$ & $\sigma_{\rm cl}$& R$_{\rm proj}$ & $\mid
\Delta_{cz} \mid$ &
References\\ & (1950) & (1950)\ \ & kms$^{-1}$ & mag & & $h^{-1}$ Mpc
& kms$^{-1}$ & $h^{-1}$ Mpc &
kms$^{-1}$ & \\ 
(1) & (2) & (3)\ \ \ \ & (4) & (5) & (6) & (7) & (8) & (9) & (10) & (11) \\
 \tableline
CGCG 385-091   &01:20:39 & -00:54:11 & \ 8269 &  14.82   & A194       & 1.042 & 408 &1.188 &    2889  & JFKa\\
NGC 0533       &01:22:57 & 01:29:57 & \ 5544 & 13.44 & A194       &1.042 & 408 & 3.018 &   \  164  & JFKa\\
UGC 01269      &01:46:11 & 34:44:05 & \ 3848 & 15.43 & A262       & 1.130& 450 &11.340 &   1134  & S \\ 
UGC 01837      &02:19:49 & 42:47:06 & \ 6582 & 14.63 & A347       & 1.000& 768 &1.357 &    1063   &  S \\
UGC 01841      &02:20:02 & 42:45:55 & \ 6373 & 14.29 & A347       & 1.000 & 768 &1.328 &     854  &  S \\
PER 195     &03:15:48 & 40:54:04 & \ 8342 & 14.20 & Perseus    & 1.110 & 550 &0.382 &    3375   & D, S\\
NGC 1705    &04:53:06 &-53:26:30 & \ \ 597 & 13.06 & Doradus   & 1.300& 245 &1.653 &    \ 517   & MDL\\
$[$WFC91$]$ 101  &13:51:06 &-29:34:59 & \ 6923 & 14.75 & Klemola27      & 0.670 & 382 &1.067 &    2312   & W\\
$[$WFC91$]$ 039  &13:59:18 &-32:55:44 & 10630 & 15.47 & AS753       & 0.979 & 401 &0.220 &    6463 & JFKa, W\\ 
$[$WFC91$]$ 056 &14:00:18 &-33:46:56 & \ 2718 & 14.37 & AS753      &0.979 & 401 &0.282 &   1449  & JFKa, W\\
ESO 384G037     &14:00:38 &-33:50:02 & \ 5723 & 14.78 & AS753      & 0.979& 401 &0.345 &    1556  & JFKa, W\\
D69        &23:44:37 &-28:17:42 & 10071 & 16.88 & Klemola44       & 1.034 & 375 &0.134 &    1614   & JFKa, LC\\
D82        &23:45:20 &-28:13:54 & 10688 & 17.12 & Klemola44       & 1.034 & 375 &0.261 &    2231   & LC\\
D39        &23:45:43 &-28:27:24 & 10554 & 16.30 & Klemola44       & 1.034 & 375 &0.305 &    2097  & JFKa, LC\\
D65        &23:45:47 &-28:21:10 & 10230 & 15.00 & Klemola44      & 1.034 & 375 &0.298 &    1773   & JFKa, LC\\
 \tableline
\end{tabular}
\end{center}
\footnotesize
References. --- D: Dressler et al. (1987);
JFKa: J{\o}rgensen et al. (1995a);
LC: Lucey \& Carter (1988); 
MDL: groups from Maia et al. (1989);
S: Smith et al. (1997); and
W: Willmer et al. (1991).
\end{table}

\clearpage

\normalsize
\begin{table}
\begin{center}
\caption{External comparisons of spectroscopic parameters}
\vspace*{10truemm}
\small
\begin{tabular}{lrccrcc}
 \tableline  \tableline
\label{tab:corrsp}
Sources & N$_{c}$ &  $\Delta \log{\sigma}$ & $\sigma_{\Delta \log{\sigma}}$ & N$_{c}$ &
 $\Delta$ Mg$_2$ & $\sigma_{\Delta {\rm Mg}_2}$ \\
(1) & (2) & (3) & (4) & (5) & (6) & (7) \\
 \tableline
LC      &  29 & 0.012$\pm 0.011$ & 0.054 & -- & -- & -- \\
7S      & 115 & 0.019$\pm 0.005$ & 0.055 & 82 & 0.004$\pm 0.002$ & 0.021 \\
D       &  25 & 0.025$\pm 0.012$ & 0.055 & 23 & 0.005$\pm 0.005$ & 0.023 \\
JFKb    &  64 & 0.021$\pm 0.008$ & 0.059 & 39 & 0.004$\pm 0.004$ & 0.023 \\
Lc      &   7 & 0.009$\pm 0.011$ & 0.030 &  7 & 0.003$\pm 0.006$ & 0.019 \\
S       &   8 & 0.011$\pm 0.014$ & 0.043 &  5 & 0.002$\pm 0.007$ & 0.021 \\
 \tableline
\end{tabular}
\end{center}
\footnotesize
References. ---
As in Table~\ref{tab:speobcl}.
\end{table}

\clearpage

\begin{table}
\begin{center}
\caption{Spectroscopy: sources of the cluster sample}
\vspace*{10truemm}
\small
\begin{tabular}{lcccccc}
 \tableline  \tableline
\label{tab:speobcl}
Source & $N_{\sigma}$(meas) & $N_\sigma$(gal) & $N_\sigma$(sing) &
$N_{{\rm Mg}_2}$(meas) & $N_{{\rm Mg}_2}$(gal) & $N_{{\rm Mg}_2}$(sing)\\
(1) &  (2) & (3) & (4) & (5) & (6) & (7)\\
 \tableline
Our       &  338 & 229 & 91 &  333 & 225 & 105 \\
\\[-2mm]
LC        & &  89 &  44 & & --- &-- \\
7S        & & 208 &  58 & & 155 &64 \\   
D         & &  81 &  31 & &\ 53 &26 \\
JFKb      & & 157 &  75 & & 129 &91 \\
Lc        & &\ 85 & \ 1 & &\ 65 &45 \\
S         & & 110 &  65 & &\ 98 &71 \\
 \tableline
\end{tabular}
\end{center}
\footnotesize
\hspace{1.5truecm}
References. ---
JFKb: J{\o}rgensen et al. (1995b);
other references are as in Table~\ref{tab:clusters}.
\end{table}

\clearpage

\normalsize
\begin{table}
\begin{center}
\caption{External comparisons of photometric parameters}
\vspace*{10truemm}
\small
\begin{tabular}{lcccccccccc}
 \tableline  \tableline
\label{tab:corrph}
Sources & N$_{c}$ &$\Delta \log{d_n}$ &
$\sigma_{\log{d_n}}$\\
(1) & (2) & (3) & (4) \\
 \tableline
LC  & 35 & 0.005$\pm 0.004$ & 0.023 \\
7S      & 162 & 0.006$\pm 0.002$ & 0.024 \\
D       & 31  & 0.010$\pm 0.003$ & 0.016 \\
JFKa & 124 & -0.004$\pm 0.002$ & 0.013  \\
Lc  & 13 & -0.007$\pm 0.004$ & 0.016 \\
S   & 14 & -0.001$\pm 0.005$ & 0.019 \\
\tableline
\end{tabular}
\end{center}
\footnotesize
\hspace{1.5truecm}
References. ---
As in Table~\ref{tab:phoobcl}.
\end{table}

\clearpage

\begin{table}
\begin{center}
\caption{Photometry: sources of the cluster sample}
\vspace*{10truemm}
\small
\begin{tabular}{lcccc}
 \tableline  \tableline
\label{tab:phoobcl}
Source & Filter & $N_{d_n}$(meas) & $N_{d_n}$(gal) & $N_{d_n}$(sing) \\
(1) &  (2) & (3) & (4) & (5) \\
 \tableline
Our            & R &  508 &  348 & 117  \\              
\\[-2mm]				      
LC             & V & &\ 77 &  19  \\ 
7S             & B & & 199 &  42  \\  
D              & B & &\ 63 & \ 9  \\ 
JFKa           & Gunn-r & &197 & 59 \\
Lc             & V & &\ 86 & \ 0 \\
S              & R & &113 &  65 \\ 
 \tableline
\end{tabular}
\end{center}
\footnotesize
\hspace{1.5truecm} 
References. ---
JFKa: J{\o}rgensen et al. (1995a);
other references are as in Table~\ref{tab:speobcl}.
\end{table}

\clearpage

\begin{figure}
\centering
\mbox{\psfig{figure=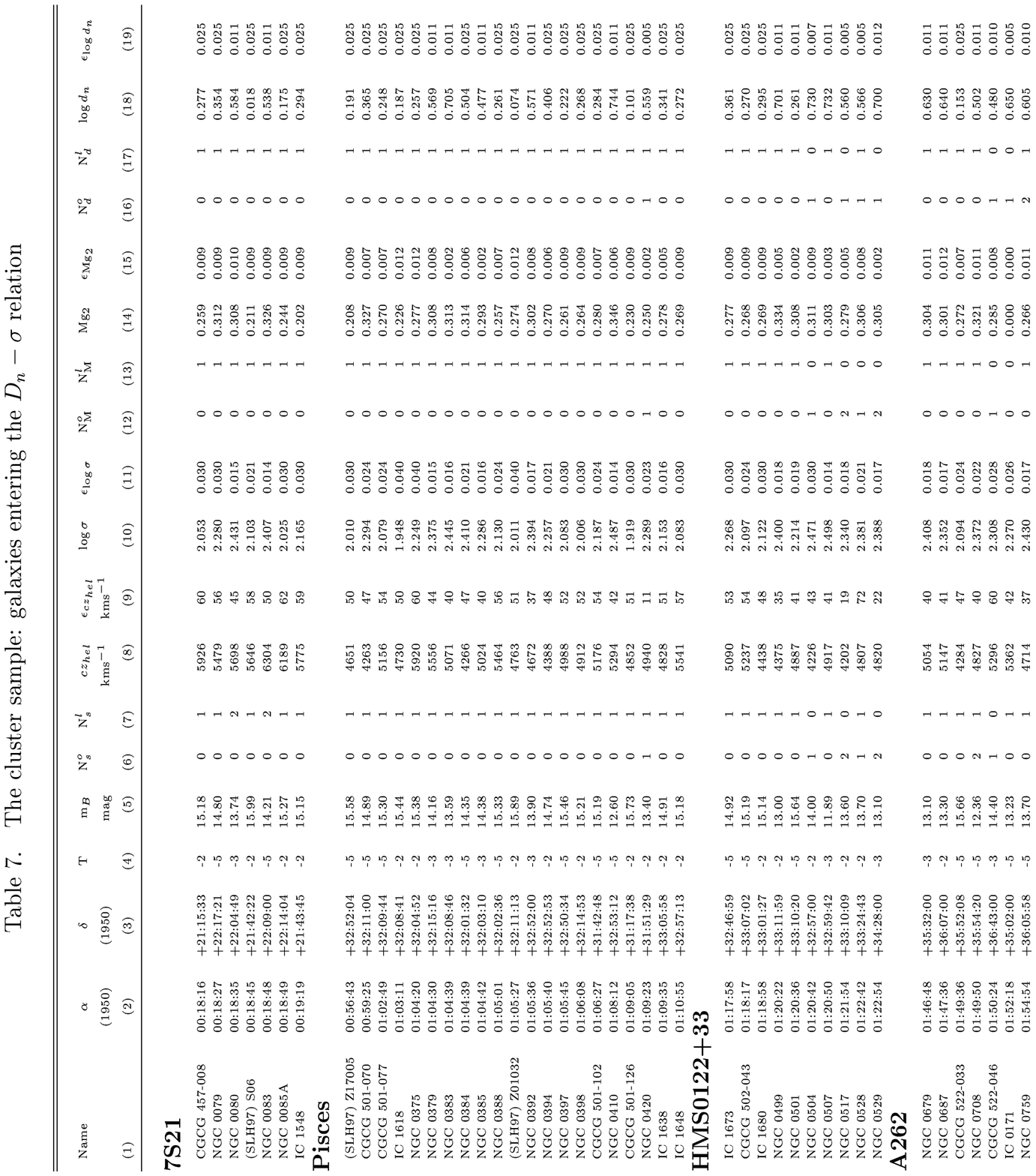,height=12truecm,bbllx=2truecm,bblly=9truecm,bburx=19truecm,bbury=21truecm}}
\end{figure}

\clearpage

\begin{figure}
\centering
\mbox{\psfig{figure=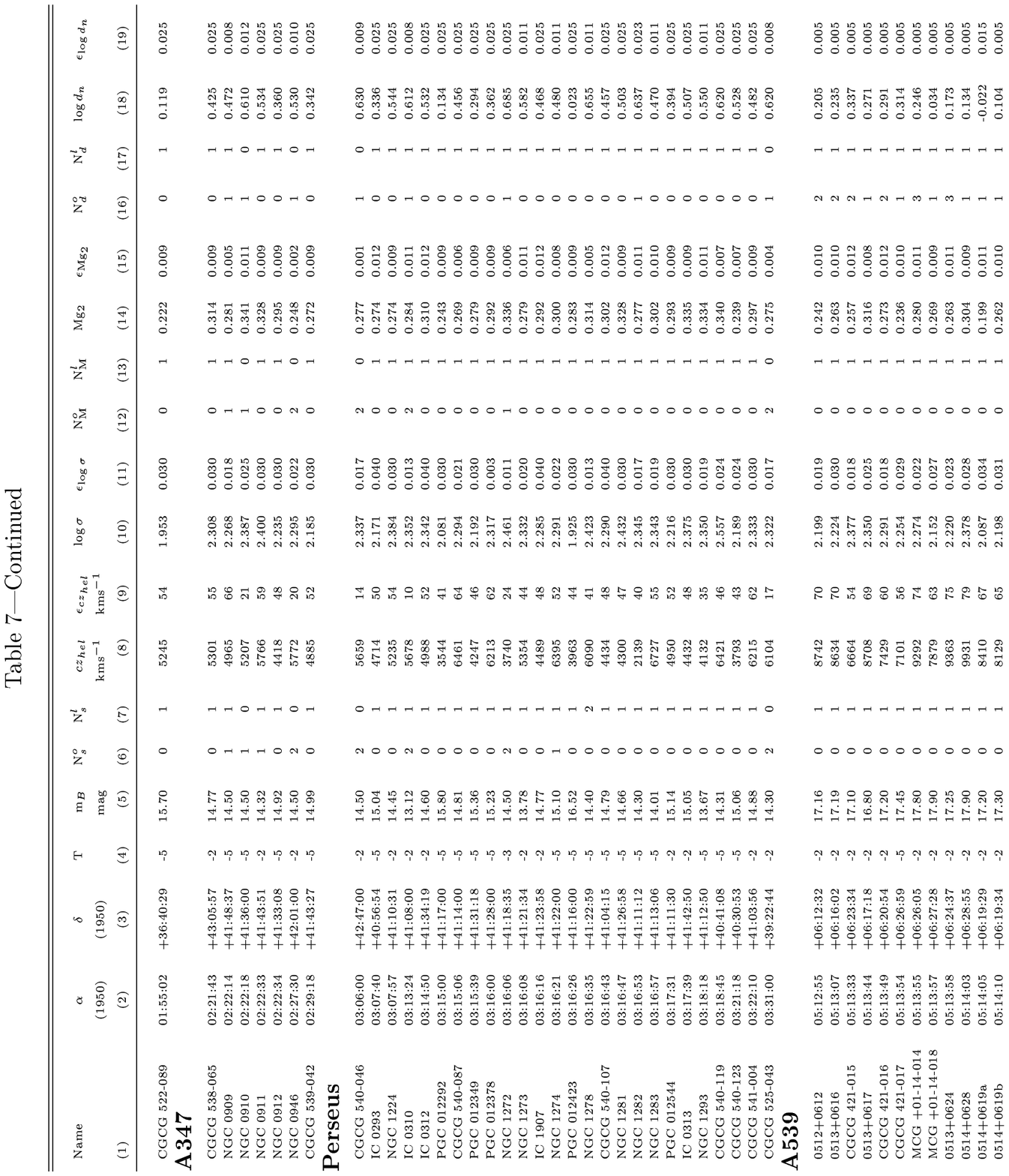,height=12truecm,bbllx=2truecm,bblly=9truecm,bburx=19truecm,bbury=21truecm}}
\end{figure}

\clearpage

\begin{figure}
\centering
\mbox{\psfig{figure=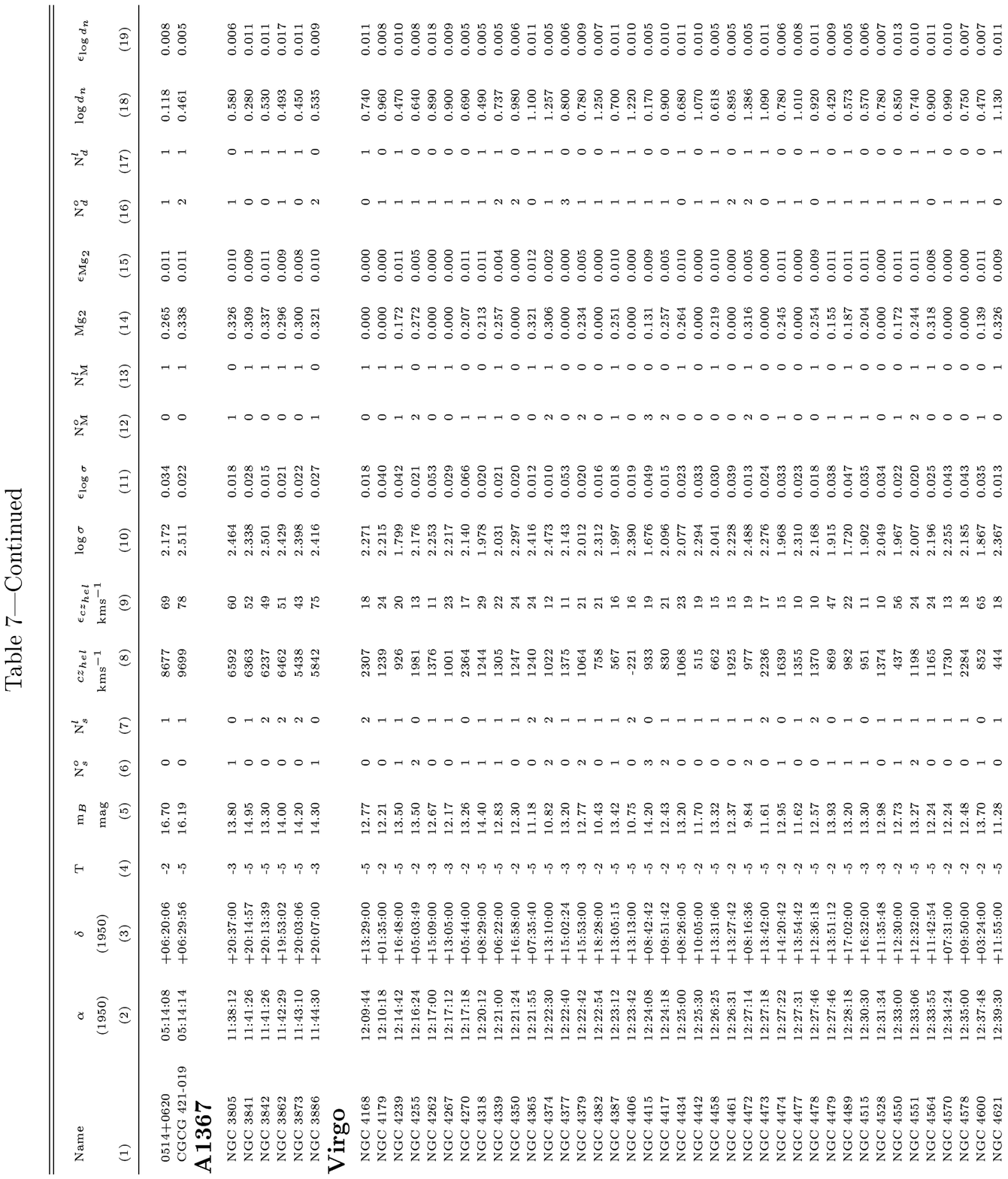,height=12truecm,bbllx=2truecm,bblly=9truecm,bburx=19truecm,bbury=21truecm}}
\end{figure}

\begin{figure}
\centering
\mbox{\psfig{figure=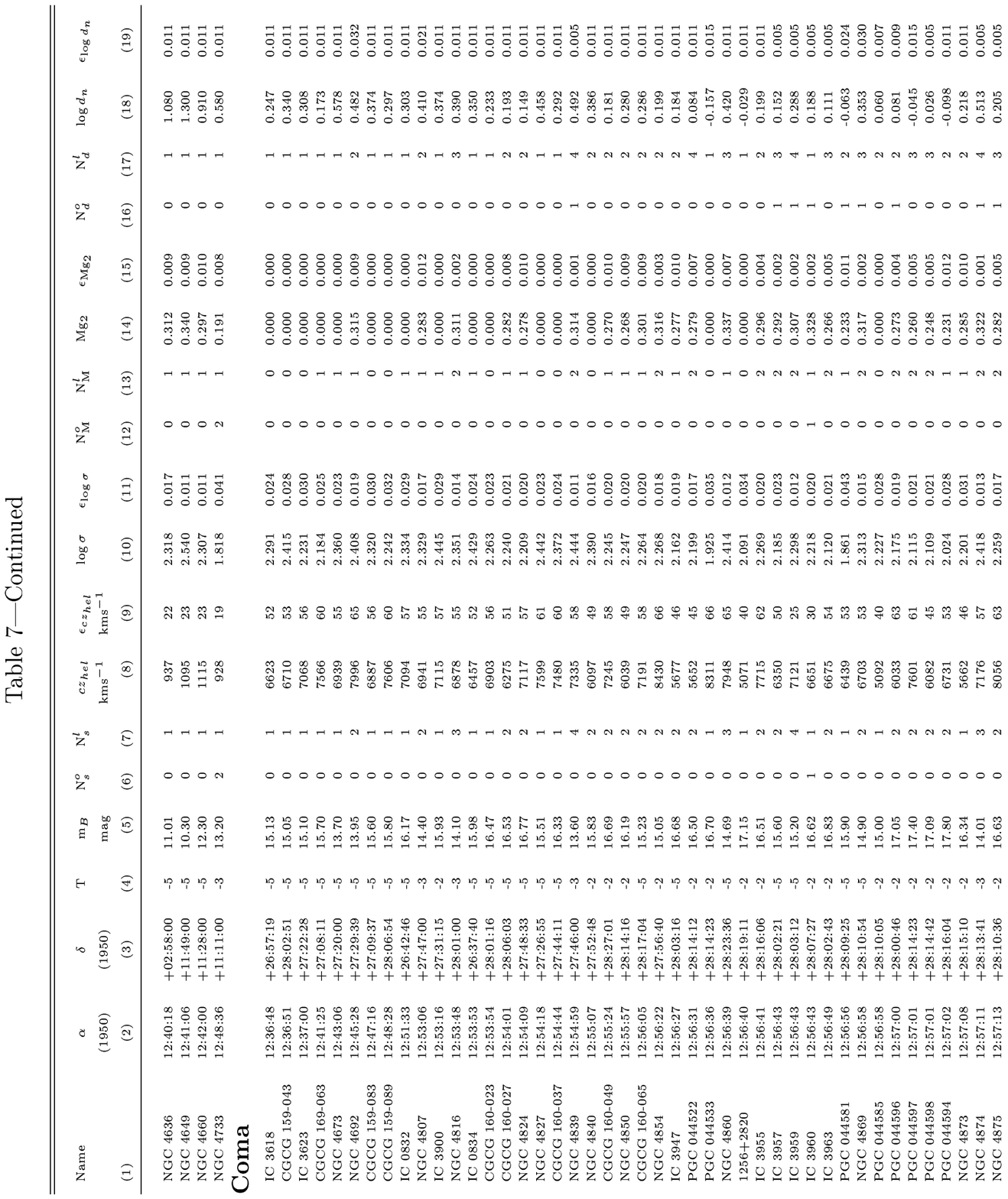,height=12truecm,bbllx=2truecm,bblly=9truecm,bburx=19truecm,bbury=21truecm}}
\end{figure}

\begin{figure}
\centering
\mbox{\psfig{figure=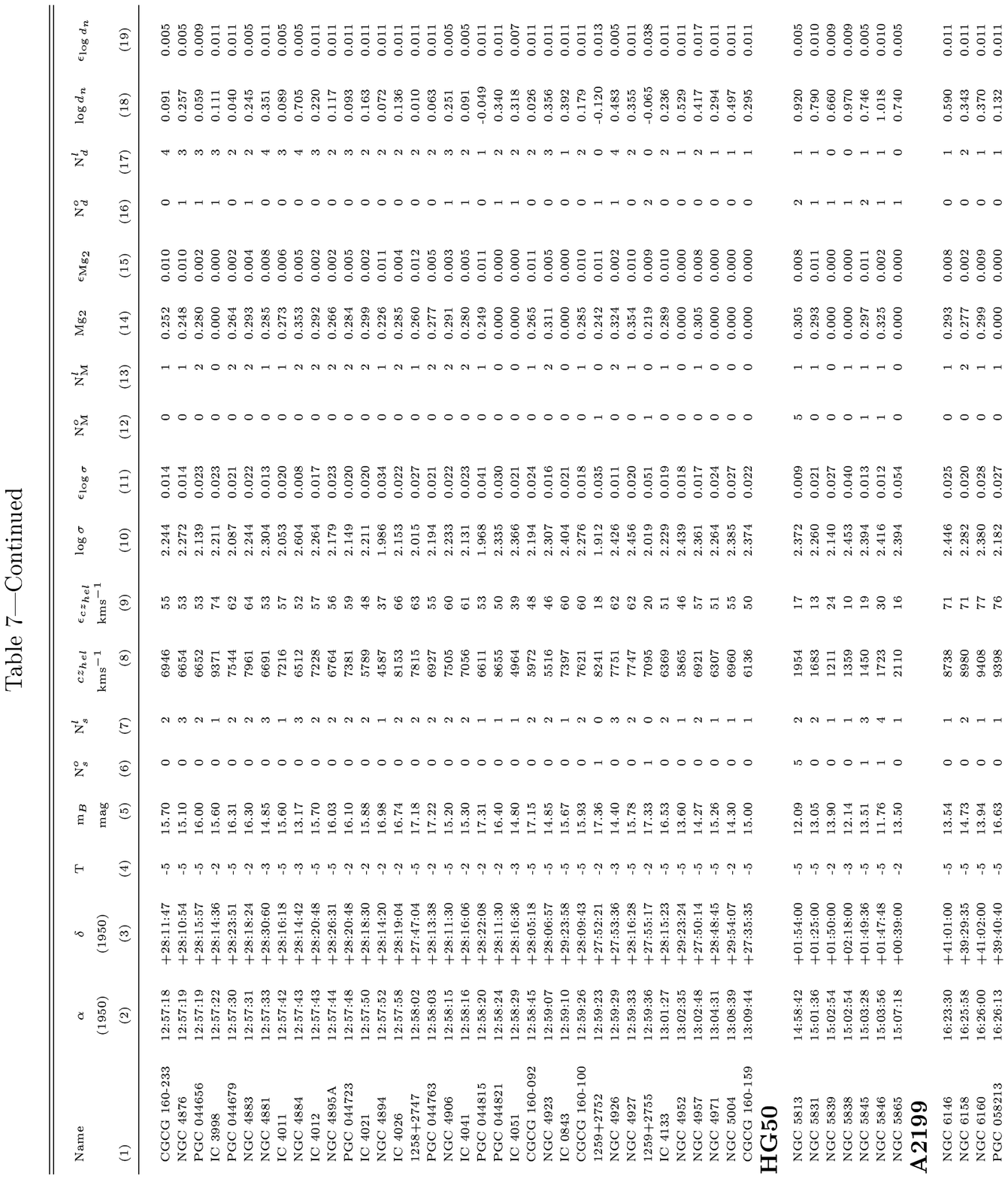,height=12truecm,bbllx=2truecm,bblly=9truecm,bburx=19truecm,bbury=21truecm}}
\end{figure}

\begin{figure}
\centering
\mbox{\psfig{figure=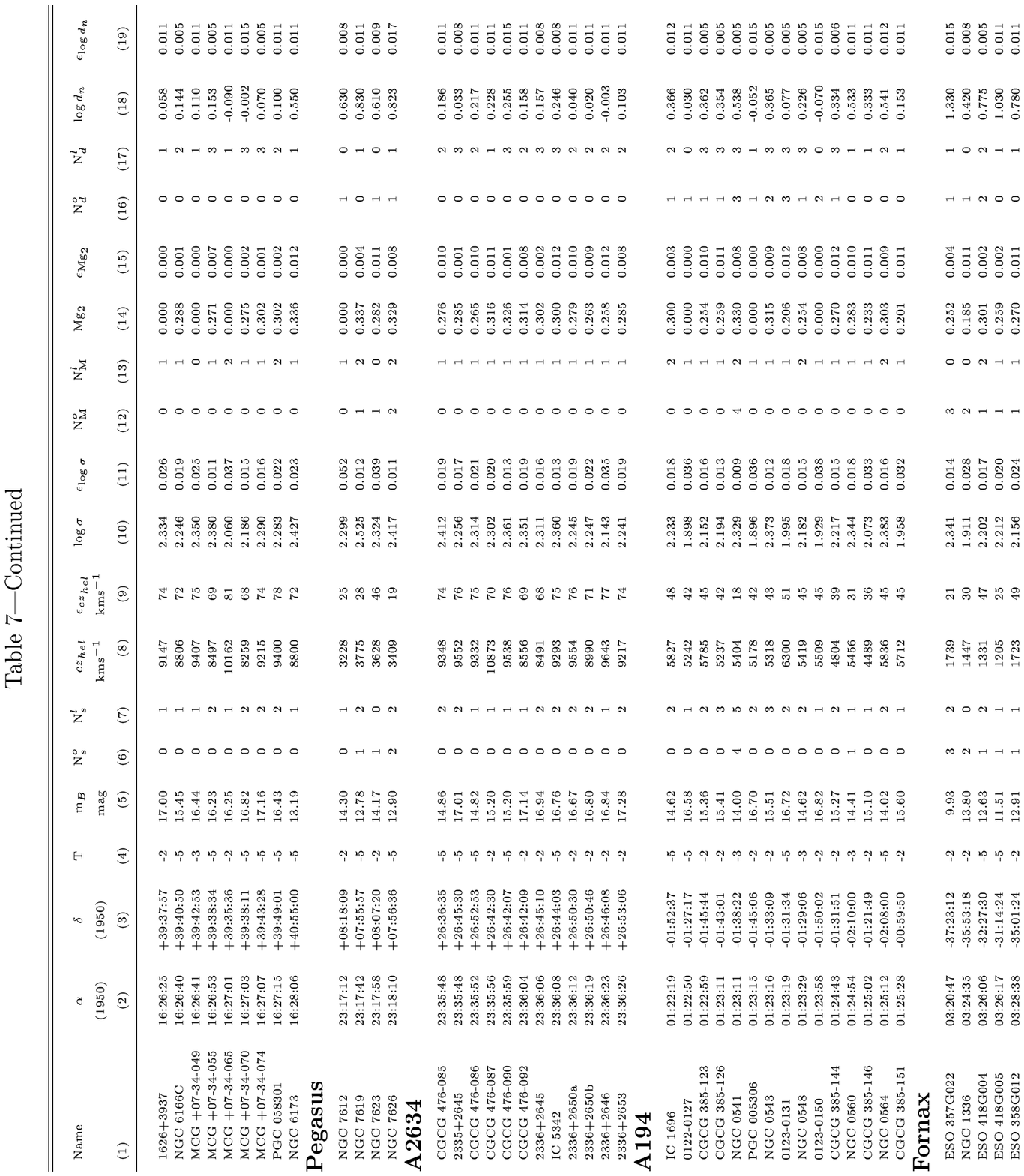,height=12truecm,bbllx=2truecm,bblly=9truecm,bburx=19truecm,bbury=21truecm}}
\end{figure}

\begin{figure}
\centering
\mbox{\psfig{figure=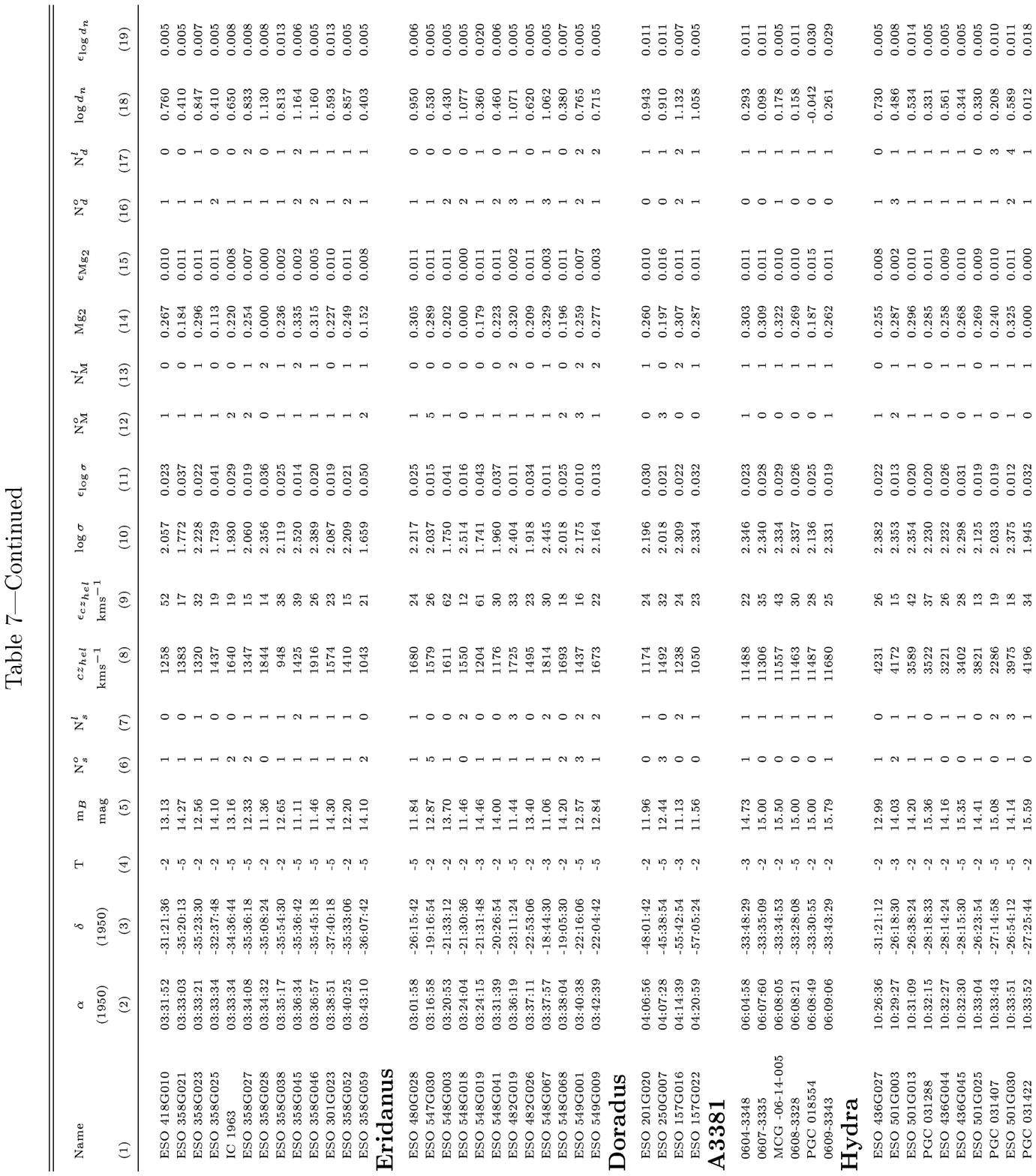,height=12truecm,bbllx=2truecm,bblly=9truecm,bburx=19truecm,bbury=21truecm}}
\end{figure}

\begin{figure}
\centering
\mbox{\psfig{figure=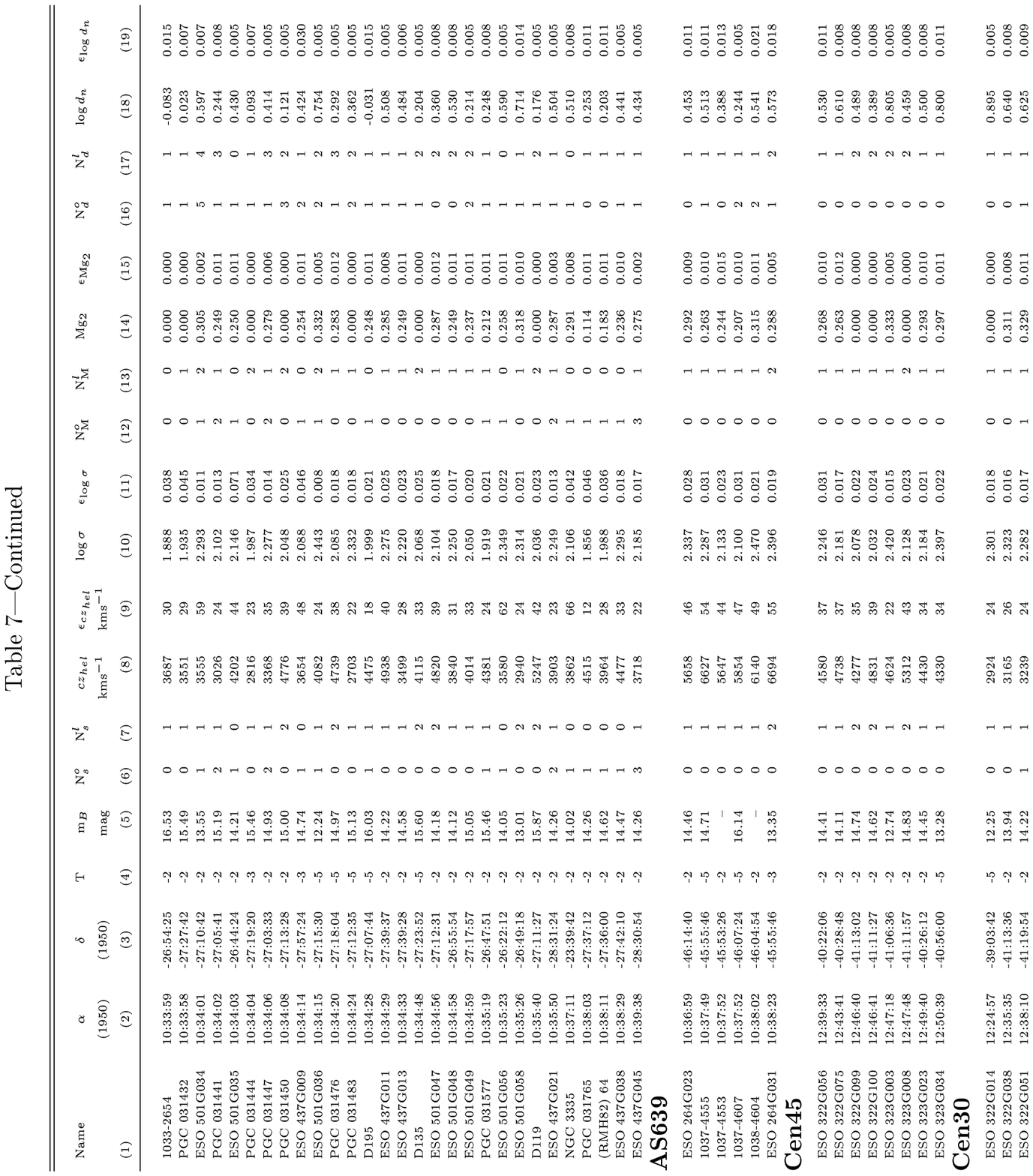,height=12truecm,bbllx=2truecm,bblly=9truecm,bburx=19truecm,bbury=21truecm}}
\end{figure}

\begin{figure}
\centering
\mbox{\psfig{figure=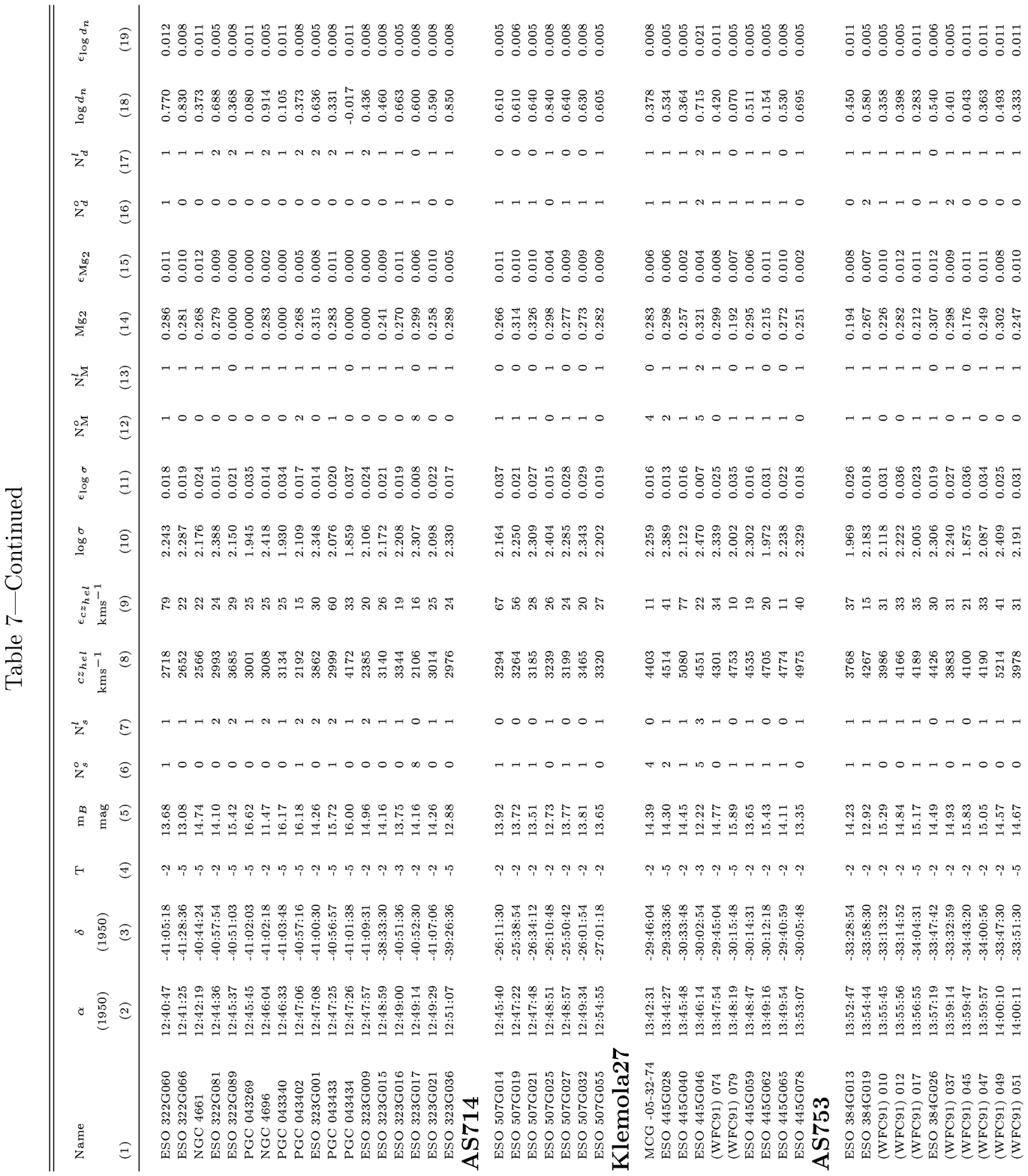,height=12truecm,bbllx=2truecm,bblly=9truecm,bburx=19truecm,bbury=21truecm}}
\end{figure}

\begin{figure}
\centering
\mbox{\psfig{figure=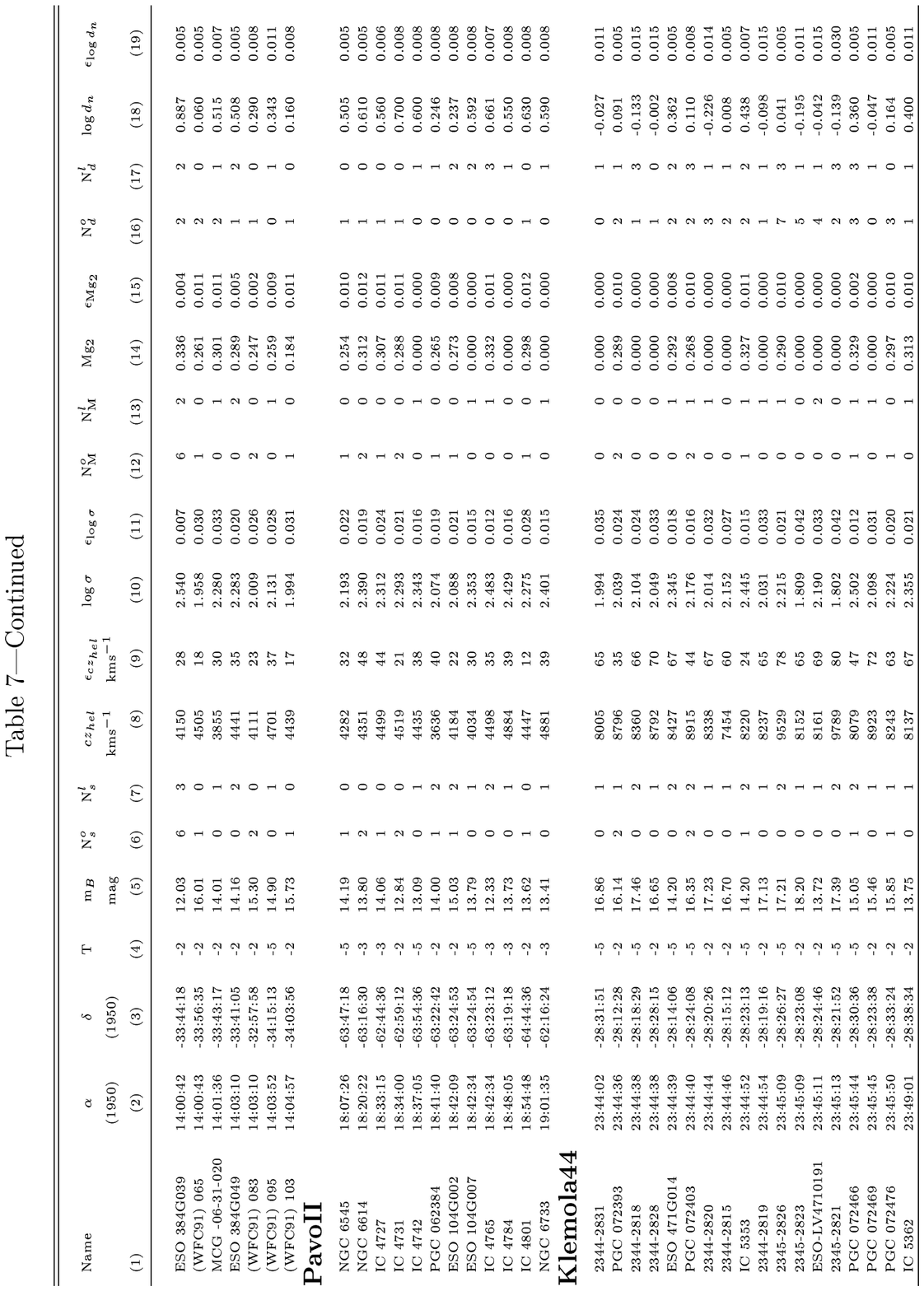,height=12truecm,bbllx=2truecm,bblly=9truecm,bburx=19truecm,bbury=21truecm}}
\end{figure}

\clearpage

\begin{figure}
\centering
\mbox{\psfig{figure=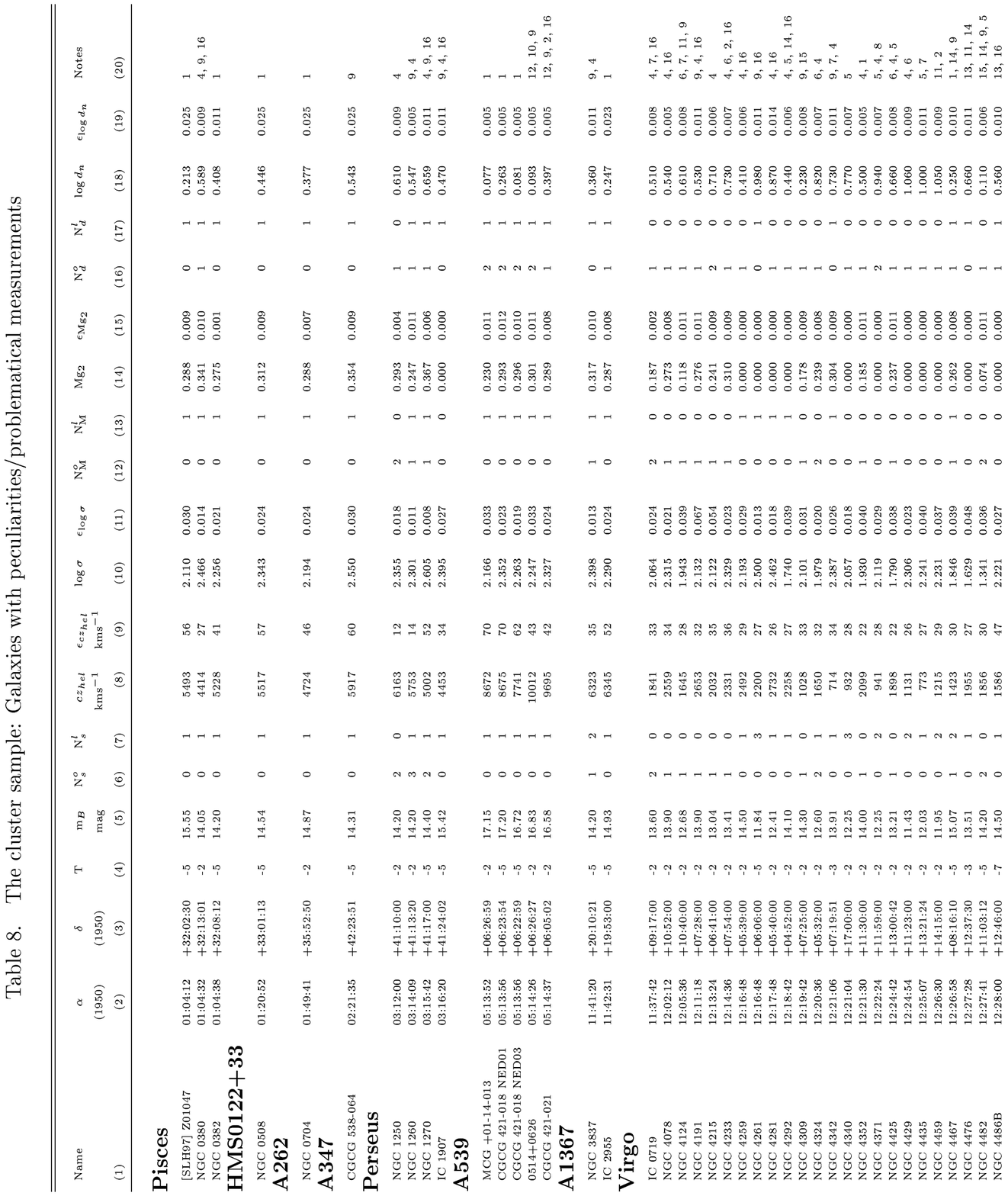,height=12truecm,bbllx=2.5truecm,bblly=9truecm,bburx=19truecm,bbury=21truecm}}
\end{figure}

\clearpage

\begin{figure}
\centering
\mbox{\psfig{figure=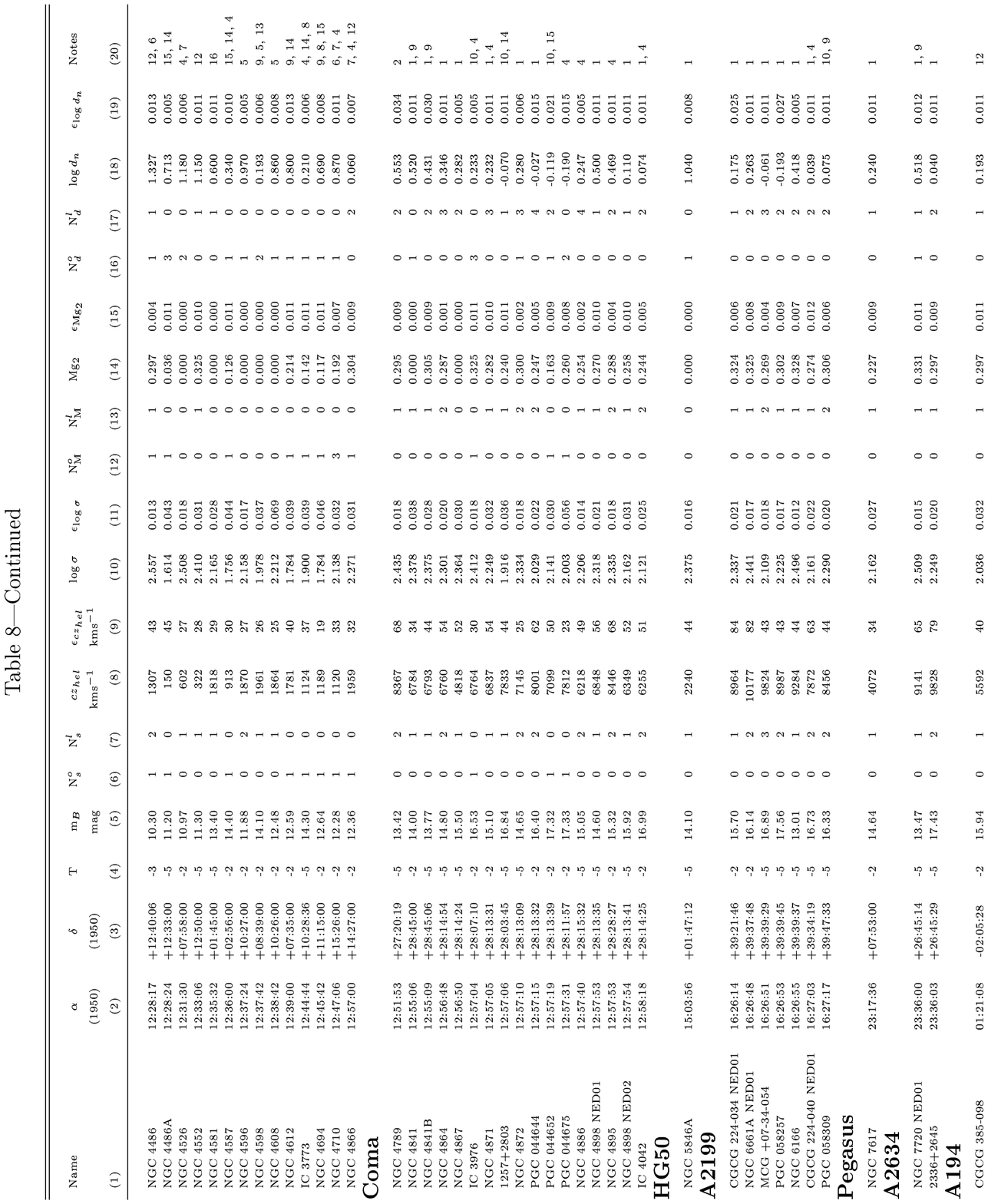,height=12truecm,bbllx=2truecm,bblly=9truecm,bburx=19truecm,bbury=21truecm}}
\end{figure}

\clearpage

\begin{figure}
\centering
\mbox{\psfig{figure=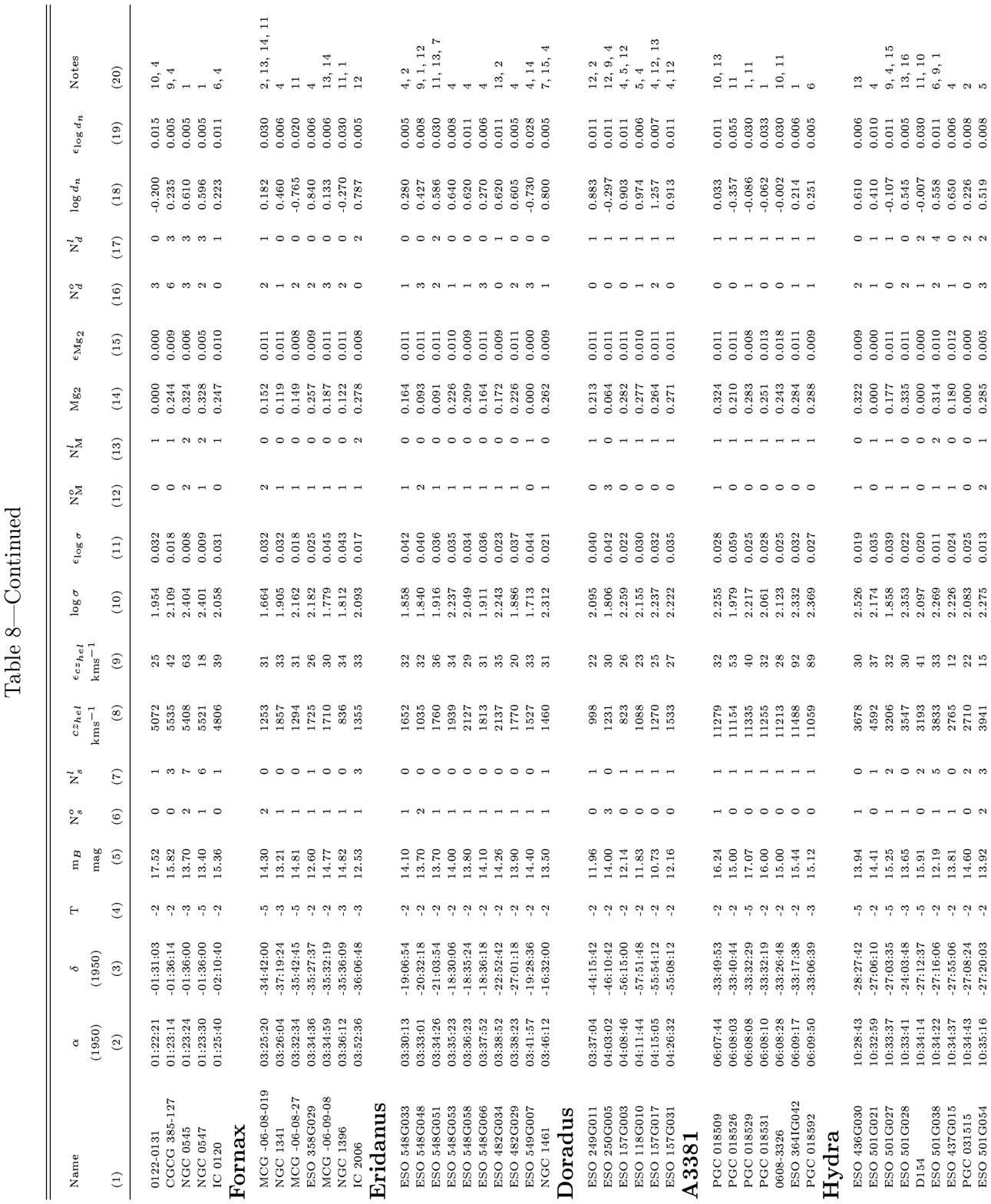,height=12truecm,bbllx=2truecm,bblly=9truecm,bburx=19truecm,bbury=21truecm}}
\end{figure}

\begin{figure}
\centering
\mbox{\psfig{figure=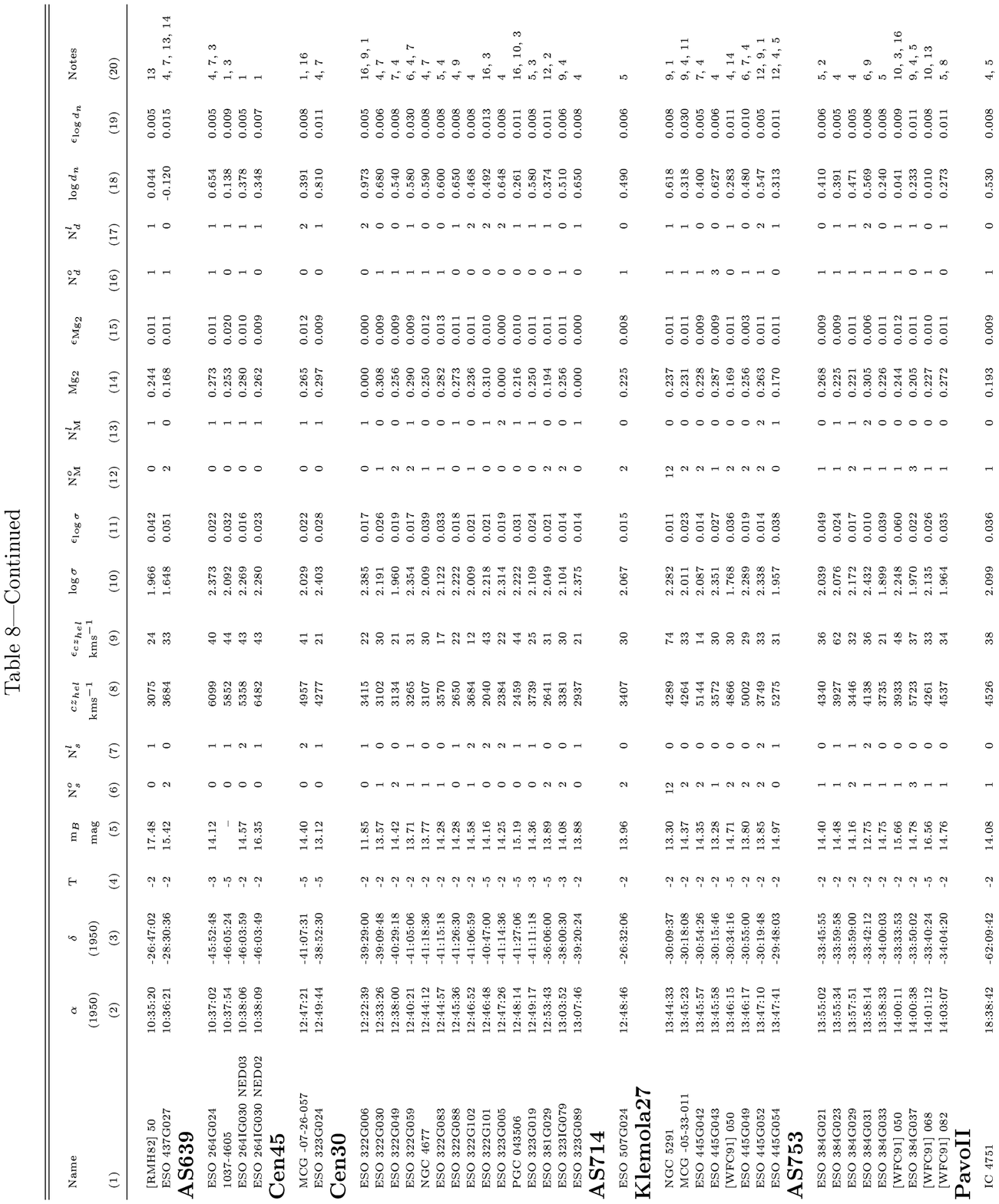,height=12truecm,bbllx=2truecm,bblly=9truecm,bburx=19truecm,bbury=21truecm}}
\end{figure}

\begin{figure}
\centering
\mbox{\psfig{figure=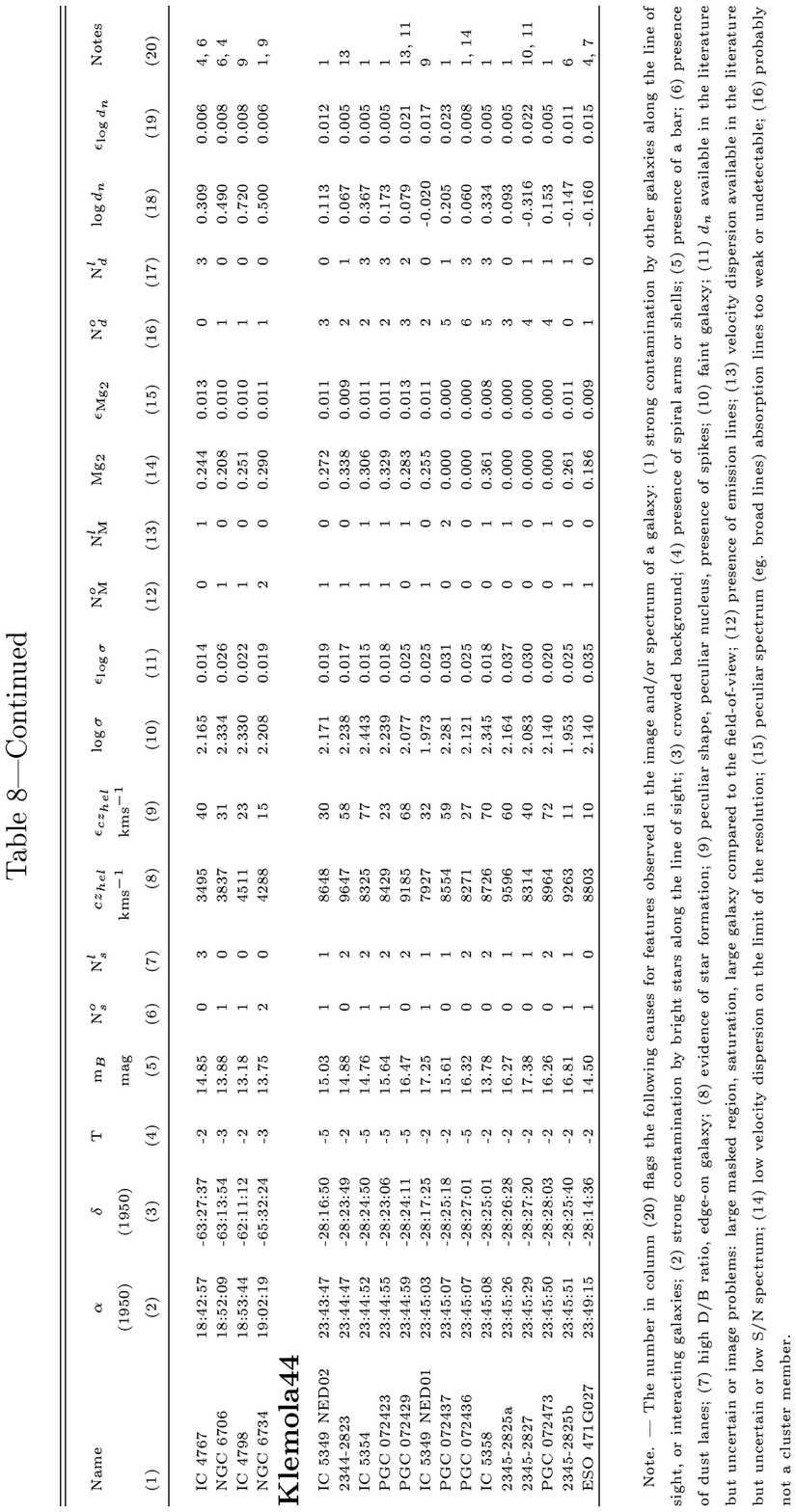,height=12truecm,bbllx=2truecm,bblly=9truecm,bburx=19truecm,bbury=21truecm}}
\end{figure}

\end{document}